\documentclass[10pt,journal,compsoc]{IEEEtran}
\usepackage{hyperref}
\usepackage{bm}
\usepackage{amssymb}
\usepackage{booktabs}
\usepackage{multirow}
\usepackage{xcolor}
\usepackage{listings}
\ifCLASSOPTIONcompsoc
  \usepackage[nocompress]{cite}
\else
  \usepackage{cite}
\fi
\ifCLASSINFOpdf
  \usepackage[pdftex]{graphicx}
  \graphicspath{{./figures/}}
  \DeclareGraphicsExtensions{.pdf,.jpeg,.jpg,.png}
\else
\fi
\usepackage{amsmath}
\ifCLASSOPTIONcompsoc
  \usepackage[caption=false,font=footnotesize,labelfont=sf,textfont=sf]{subfig}
\else
  \usepackage[caption=false,font=footnotesize]{subfig}
\fi
\begin{document}
\title{PassSum: Leveraging Paths of Abstract\\ Syntax Trees and Self-Supervision for\\ Code Summarization}

\author{Changan Niu, Chuanyi Li, Vincent Ng, Jidong Ge, Liguo Huang, Bin Luo
    \IEEEcompsocitemizethanks{
        \IEEEcompsocthanksitem C. Niu, C. Li, J. Ge and B. Luo are with the State Key Laboratory for Novel Software Technology at Nanjing University, Software Institute, Nanjing University, 22 Hankou Road, Nanjing, China.\protect\\
        E-mails: niu.ca@outlook.com, \{lcy,gjd,luobin\}@nju.edu.cn.
        \IEEEcompsocthanksitem V. Ng is with the Human Language Technology Research Institute, University of Texas at Dallas, Richardson, TX 75083-0688.\protect\\
        E-mail: vince@hlt.utdallas.edu.
        \IEEEcompsocthanksitem L. Huang is with the Dept. of Computer Science, Southern Methodist University, Dallas, TX 75275-0122.\protect\\
        E-mail: lghuang@lyle.smu.edu.
        \IEEEcompsocthanksitem Corresponding Author: Chuanyi Li.
    }
    \thanks{Manuscript received XXX XX, 202X; revised XXX XX, 202X.}
}


\markboth{}%,~Vol.~14, No.~8, August~2015}%
{Niu \MakeLowercase{\textit{et al.}}: PassSum: Leveraging Paths of Abstract Syntax Trees and Self-Supervision for Code Summarization}



\IEEEtitleabstractindextext{%
\begin{abstract}
Code summarization is to provide a high-level comment for a code snippet that typically describes the function and intent of the given code. Recent years have seen the successful application of data-driven code summarization. To improve the performance of the model, numerous approaches use Abstract Syntax Trees (ASTs) to represent the structural information of the code, which is considered by most researchers to be the main factor that distinguishes code from natural language. Then, such data-driven methods are trained on large-scale labeled datasets to obtain a model with strong generalization capabilities that can be applied to new examples. Nevertheless, we argue that state-of-the-art approaches suffer from two key weaknesses: (1) inefficient encoding of ASTs; (2) reliance on a large labeled corpus for model training. As a result, such drawbacks lead to (1) oversized model, slow training, information loss and instability; (2) inability to be applied to programming languages with only a small amount of labeled data. In light of these weaknesses, we propose PassSum, a code summarization approach that addresses the aforementioned weaknesses via (1) a novel input representation which contains an efficient AST encoding method; (2) introducing three pre-training objectives and pre-training our model with a large amount of (easy-to-obtain) unlabeled data under the guidance of self-supervised learning. Experimental results on code summarization for Java, Python and Ruby methods demonstrate the superiority of PassSum to state-of-the-art methods. Further experiments demonstrate that the input representation we use has both temporal and spatial advantages in addition to performance leadership. In addition, pre-training is also shown to make the model more generalizable with less labeled data, and also to speed up the convergence of the model during training.
\end{abstract}

\begin{IEEEkeywords}
Code Summarization, Abstract Syntax Tree, Transformer, Self-Supervised Pre-Training.
\end{IEEEkeywords}}

\maketitle

\IEEEdisplaynontitleabstractindextext

% \fi
\IEEEpeerreviewmaketitle

\IEEEraisesectionheading{\section{Introduction}\label{sec:introduction}}

% for IEEE Computer Society journal papers produced under \LaTeX\ using
% IEEEtran.cls version 1.8b and later.
% % You must have at least 2 lines in the paragraph with the drop letter
% % (should never be an issue)
% I wish you the best of success.

 

% \subsection{Subsection Heading Here}
% Subsection text here.

% % needed in second column of first page if using \IEEEpubid
% %\IEEEpubidadjcol

\begin{figure}[t]
    \centering
    \includegraphics[width=0.9\linewidth]{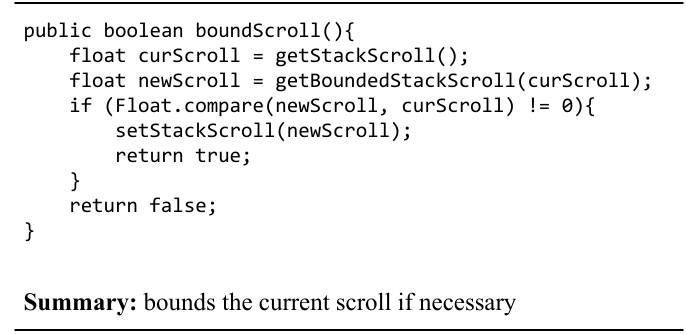}
    \caption{An example of code summarization.}
    \label{figure:summarization_example}
\end{figure}

\IEEEPARstart{C}{ode} summarization, also known as the process of code summary or code comment generation, is the task of automatically generating a natural language description for a piece of source code to summarize the overall actions of the code snippet accurately, adequately and concisely~\cite{sridhara2010towards}. Work in this area of research has generally focused on generating a (typically short) natural language comment from a given method in a piece of source code. Consider the example in \figurename~\ref{figure:summarization_example}.
Given the Java method \texttt{boundScroll} as the input, a code summarization system should generate the summary ``bounds the current scroll if necessary''.

Code comments play an important role in software quality assurance by making it easier to understand, reuse and maintain. Early studies show that some artifacts can improve the readability of programs or procedures, like data structure descriptions~\cite{shneiderman1982control}, documentations~\cite{norcio1982indentation} and comments~\cite{tenny1985procedures}. In order to assess the level whether impact that each one of the 25 proposed readability features may have, positively or negatively on the overall code readability, Tashtoush et al.~\cite{tashtoush2013impact} distribute a survey to a random number of expert programmers. They find that the comment ranks third in terms of positively impacting features, after meaningful names and consistency. Souza et al.~\cite{de2005study} conduct two surveys among software maintainers to identify the most important documentation artifacts. In the first survey, the maintainers are asked what artifacts they think important or not. In the second one, they are asked what artifacts they had actually used. Results show that source code and comments are the most important artifacts to understand a system to be maintained. In addition, Marin~\cite{marin2005motivates} finds that when programmers have a piece of code in front of them that is thoroughly commented, they will write comments for the newly added code.

Another clue is that if the program is not well-commented, the other developers would spend almost about 58\% of their time on program comprehension before coding according to~\cite{xia2017measuring}. Fluri et al.~\cite{fluri2007code} conduct three experiments on three open source software systems (i.e., ArgoUML, Azureus, and JDT Core) to investigate the co-evolution relation between source code and its associated comments. Their findings show that code and comments are rarely co-evolve and newly added code is barely commented. In addition, Forward et al.~\cite{forward2002relevance} learn the fact that documentation is almost always out-dated and inconsistent. Therefore, code summarization proves to be a meaningful task.

In order to summarize the functionality of a piece of code into a natural language, earlier code summarization methods mainly use the information retrieval (IR) method. Data-driven approaches have taken over this area of research since the first neural network-based approach, named CODE-NN~\cite{iyer2016summarizing}, is applied to code summarization. Since the neural network approach is data-sensitive, a large-scale and high-quality dataset is always a key factor of the performance of these methods. However, building such dataset requires excessive costs because annotating source code data requires significant programming knowledge and expertise.

On the other side, for further exploitation of the structural information of the code, in recent years, virtually all recently developed systems take the corresponding AST as the additional or only input~\cite{alon2019code2vec,leclair2019neural,zhang2019novel,hu2020deep,leclair2020improved,lin2021improving}. Therefore, how to encode the AST becomes another critical factor of the code summarization method.

Despite recent significant progress on code summarization, these state-of-the-art approaches still suffer from some weaknesses, such as the two aforementioned aspects. First, deep neural networks require existing methods to rely on vast volumes of high-quality supervised data, which is currently unavailable in the literature. The second is that the current AST encoding method is inefficient, as we will see in the next section.

In order to address the two shortcommings, in this paper, we propose PassSum (abbreviation of leveraging \textbf{Pa}ths of ASTs and \textbf{s}elf-\textbf{s}upervision for code \textbf{Sum}marization), a neural code summarizer built on Transformer~\cite{vaswani2017attention}, to address these two weaknesses. First, in light of the inefficient encoding of ASTs, we utilize a code2vec-like encoding. Rather than using path-context, a set of paths from the root node to the terminal node in AST is used as one of the inputs. Instead of random sampling like code2vec, we use the self-attention mechanism to automatically learn the relationship between the paths. This allows us to dramatically reduce the amount of the input compared to code2vec (from $P(n,2)$ to $n$) while still keeping its design intent, resulting in a novel AST encoding method we call \textit{Path Embedding}. Aside from the previously described path input, we also use the code sequence as an input to enable the model to learn both syntactical and lexical information.

Second, to improve the robustness of a code summarizer's performance in the face of a small training set, we propose to leverage \textit{unlabeled} data. In our case, unlabeled data refers to those methods in the source code that do \textit{not} have comments. Note that such methods can be obtained inexpensively and in very large quantities. The question, then, is: how can a large amount of unlabeled data be leveraged for code summarization? Motivated by large pre-trained models like BERT~\cite{devlin2019bert}, we propose to use \textit{self-supervised learning} to \textit{pre-train} a code summarizer. The term \textit{Self-supervised} indicates that the supervision employed for pre-training are automatically collected from the input and therefore no longer requires manual labeling. The idea behind pre-training is simple. Before we train a model for a target task (which in our case is code summarization), we start by pre-training it on a \textit{related} task. After pre-training, knowledge learned from the related task will be stored in the parameters of the model. Then we can continue to train the model on our target task, knowledge of the related task can be transferred to the target task, potentially improving its performance. The next question, then, is: how can we use self-supervised learning to pre-train our code summarizer on \textit{unlabeled} data? As we will see in Section~\ref{section:approach}, three pre-training objectives are introduced to help our model to learn the knowledge from completely unlabeled data.

In sum, our work makes the following contributions:
\begin{itemize}
    \item We propose a novel input representation containing an efficient AST encoding method with syntactical information and an additional input part with lexical information.
    \item To improve the robust performance of PassSum, we propose to pre-train PassSum using three pre-training objectives on a large amount of unlabeled dataset.
    \item We empirically show that PassSum outperforms state-of-the-art code summarizers, regardless of the amount of data available for training the summarizers. In addition, we conduct sufficient experiments to demonstrate the time and space efficiency of the input representation used by PassSum.
\end{itemize}

We also provide the source code and data for future study and extension\footnote{https://github.com/NougatCA/PassSum}. The remainder of this paper is organized as follows. Section~\ref{section:limitations} details two shortcomings of the current approach that we have identified. Section~\ref{section:preliminaries} illustrates some preliminary background knowledge. The details of our proposed approach are elaborated in Section~\ref{section:approach}. Section~\ref{section:evaluation} demonstrates the experimental setups, results and analysis. Threats to validity are indicated in Section~\ref{section:threats}. Section~\ref{section:related_work} reviews the related work. We conclude the paper in Section~\ref{section:conclusion}.

\section{Limitations with Current Code Summarization Techniques}
\label{section:limitations}

In this section, we first present two main drawbacks w.r.t. the encoding of ASTs and the reliance on the labeled data. Then, to address these two shortcomings, we introduce our work in this paper.

\subsection{\textcolor{red}{Inefficient Encoding of ASTs}}
\label{section:limitations_ast}

The AST contains information about the structure of the code which is not obvious in the code sequence. As a result, much of the recent work has exploited ASTs to summarize source code. However, we find that these methods still suffer from some shortcomings. For better illustration, we classify them into 4 categories and present them separately.

\textbf{(1) Graphical Neural Networks.} Some work utilizes neural networks that can take trees as input (e.g., tree- LSTMs~\cite{wan2018improving,wan2019multi,lin2021improving}, RvNNs~\cite{zhang2019novel} and GNNs~\cite{yang2021multi,leclair2020improved,liu2021retrieval}), however, these networks are shown to be inefficient to encode the graphical input. For example, when Tree-LSTMs model tree conversations as branches of the tree, there is repetition of input as many branches share nodes and there is no communication between branches during the learning process~\cite{kumar2019tree}. And for RvNNs, we argue that they are computationally expensive due to the recursive nature of the network. GNNs also require high computational cost~\cite{cheng2021extremely}, in addition, they suffer from long dependencies~\cite{liu2021eignn,rampavsek2022recipe}, over-smoothing, and over-squashing problems~\cite{Oono2020Graph,chen2020measuring,kreuzer2021rethinking,alon2021on,topping2022understanding} as well. And it is also demonstrated that GNNs are not well-suitable for handling structured data with
varying sizes and shapes~\cite{Vashishth2020Composition-based}, while AST is just such a structured data.

\textbf{(2) Splitting into subtrees.} Secondly, splitting an AST into several subtrees and encoding them separately seems like a good idea~\cite{zhang2019novel,lin2021improving}, but this may not allow for detailed correlations between the terminal nodes from the different parts. Those correlations, however, have been greatly applied in code2vec~\cite{alon2019code2vec} and code2seq~\cite{alon2018code2seq} with outstanding results.

\begin{figure}
    \centering
    \includegraphics[width=\linewidth]{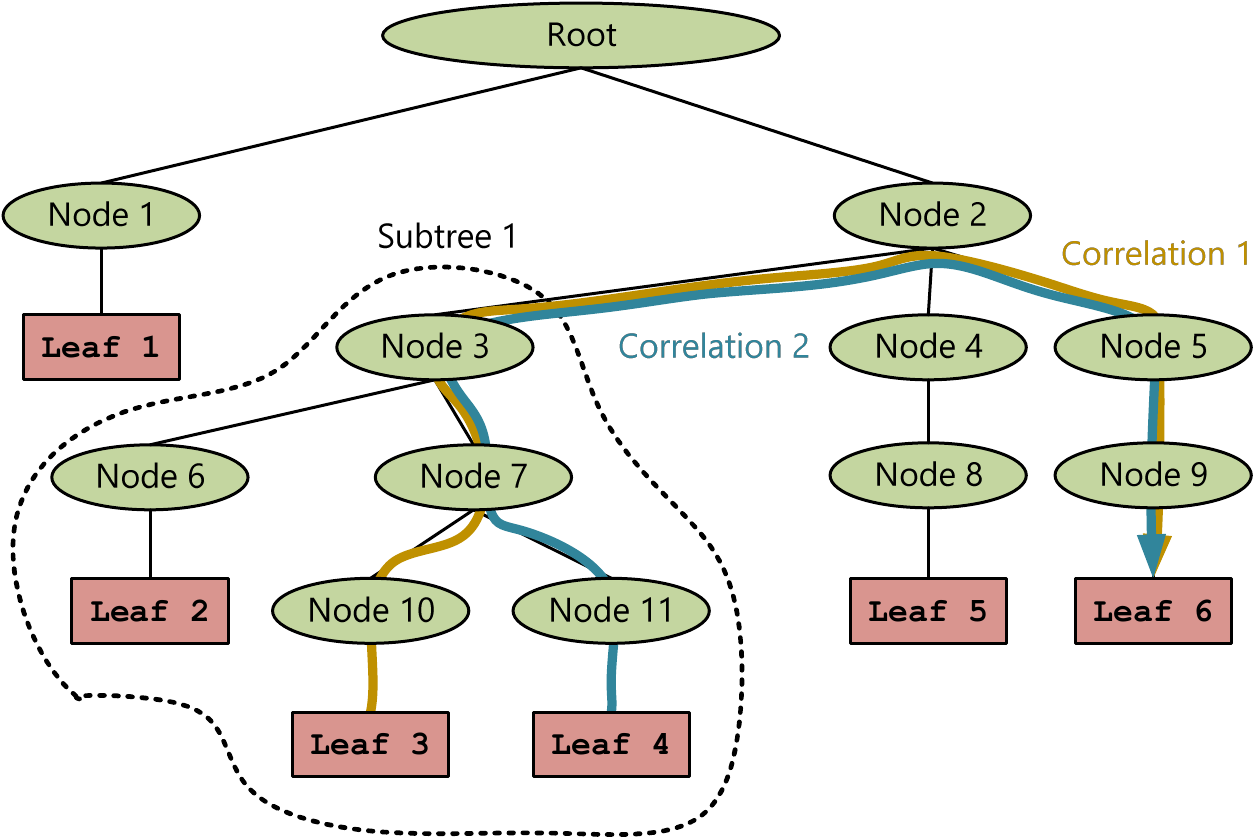}
    \caption{Two critical correlations in an AST, their starting leaves are different but both in Subtree 1.}
    \label{figure:subtree}
\end{figure}

Take \figurename~\ref{figure:subtree} as an example, assume that there are two paths in this AST that are critical to the understanding of the code, namely Correlation 1 (from Leaf 3 to Leaf 6) and Correlation 2 (from Leaf 4 to 6). If the Subtree 1 with Node 3 as the root is first encoded into a single vector, it will be difficult for the model to identify and distinguish between relation 1 and relation 2, which have the same end leaf (Leaf 6) and different starting leaves (Leaf 3 and 4, but both in Subtree 1). However, Subtree 1 is already encoded as a vector, which makes it difficult for the model to separate the two different leaves from a single vector and find the respective relationships.

\textbf{(3) Converting AST into sequence.} Meanwhile, lots of existing neural code summarizers typically convert an AST into a sequence (a process that we will henceforth refer to as \textit{linearization}) before embedding it, with a method known as Structure-based Traversal (SBT)~\cite{hu2018deep,hu2020deep,leclair2019neural,kang2019assessing,yang2021multi}. We find that the sequence encoded by SBT is so long that it is often more than three times the length of the source code. Such a long sequence may increase the training time with poor spatial efficiency, which is proven by our experiments in Section~\ref{section:rq2}.

\textbf{(4) Path-contexts.} Another linearization approach, used by code2vec~\cite{alon2019code2vec} and code2seq~\cite{alon2018code2seq}, linearizes an AST as a series of \textit{path-contexts} representing two terminal nodes and the path between them. Code2vec has been very influential and received a lot of attention since its publication. However, when the number of terminal nodes is $n$, the number of path-contexts becomes a 2-permutation of $n$: $P(n,2)=n\cdot (n-1)$. Therefore, when the number of path-contexts exceeds a predefined value $k$, code2vec randomly samples $k$ path-contexts as the input (in fact this is almost always happening). A research on code2vec, code2seq and other models~\cite{wang2021wheacha} shows that a code snippet can be reduced to very few statements (i.e., $\leq$2), for which models make the same prediction as they do for the original snippet. In other words, a small code fragment already contains the key features that models need to correctly predict the label of programs. Because of its random sampling, code2vec becomes less stable by relying on fewer features. Even a simple, natural, semantically-preserving transformation causes it to change its prediction.

\subsection{Reliance on a Large Amount of Labeled Data}
\label{section:limitations_data}

Despite achieving state-of-the-art performance, neural code summarizers, like many other neural models, need to be trained on a large amount of labeled data. In the case of the code summarization task, the labeled data is composed of a set of methods, each of which is labeled with a corresponding comment (i.e., summary). Applying existing code summarization models that rely on large-scale labeled datasets to programming languages with relatively few labeled data, the effectiveness would be lower down. However, most of the code does not have corresponding comments~\cite{fluri2007code}. Even with the commented code, there is a high probability that their comments are still out of date~\cite{fluri2007code,forward2002relevance}. Ideally, we can build code summarizers that can offer robust performance for all programming languages rather than just those where a large labeled training set is available. However, building such summarizers remains a key research challenge in neural code summarization.

By identifying these two shortcomings of current code summarization methods, in this paper, we propose to solve these two problems using path encoding and self-supervised pre-training.

\section{Preliminaries}
\label{section:preliminaries}

In this section, we will present some preliminary background knowledge about structural information of source code and self-supervised pre-training.

\subsection{Encoding Structural Information of Source Code}
\label{section:bg_structual}

Structural information is very important for the understanding of the source code. The vast majority of approaches use ASTs to represent structural information and uses many methods to make the ASTs encodable by the model. The most straightforward approach is to input the AST directly to the tree/graph neural network. Wan et al.~\cite{wan2018improving} encodes the structure content of code via AST-based LSTM. The AST-based LSTM encodes the AST from the bottom up, and its LSTM unit has multiple forget gates, which is basically a Tree-LSTM~\cite{tai2015improved}. MMAN~\cite{wan2019multi} also uses Tree-LSTM to encode the AST, and in addition, it uses Gated Graph Neural Network (GGNN)~\cite{li2015gated} to encode the Control Flow Graph (CFG) of the code, which is also considered to contain the structural information of the code. In addition, Leclair et al.~\cite{leclair2020improved} utilizes Convolutional GNNs to encode the AST. Liu et al.~\cite{liu2021retrieval} utilizes GNNs to encode local information of Code Property Graph (CPG), which is construct based on the AST.

In addition to feeding the entire AST into the network at once, much of the work is first done before slicing the AST into subtrees. ASTNN~\cite{zhang2019novel} decomposes large ASTs of code fragments into sequences of small statement trees, then obtains statement vectors by using RvNN to recursively encode statement trees. Finally ASTNN learns the vector representations of code fragments by leveraging the RNNs and the pooling layer. Lin et al.~\cite{lin2021improving} uses block-wise code splitting and employs Tree-LSTMs to encode each AST subtree.

Instead of the direct use of neural networks that can take trees as input, most methods first convert the AST into a sequence (the process of AST linearization) and then input it into a network that models the sequence (e.g., RNN and Transformer). Zhang et al.~\cite{zhang2020retrieval} and Wang et al.~\cite{wang2021code} use pre-order traversal while Svyatkovskiy et al.~\cite{svyatkovskiy2019pythia} use in-order traversal to linearize ASTs. Hu et al.~\cite{hu2018deep} propose a novel traversal method for ASTs named Structure-based Traversal (SBT), which can losslessly linearize a given AST into a sequence by traversing it. Another linearization approach proposed by Alon et al.~\cite{alon2018general} and used by code2vec~\cite{alon2019code2vec}, code2seq~\cite{alon2018code2seq} and SLM~\cite{alon2020structural}, linearizes an AST as a series of ``path-contexts'' representing two terminal nodes and the path between them.

In this work, we do not use any of the methods described above; instead, we use a series of paths from the root node to each terminal node. By using this method, we  shorten the length of the AST after serialization, reduce information loss, and speed up the training speed and space occupation of the model. The experimental results in Section~\ref{section:evaluation} demonstrate the advantages of this method.

\subsection{Self-supervised Pre-training}
\label{section:bg_pre_train}

Pre-Training has revolutionized the way computational models are trained in the NLP community~\cite{devlin2019bert,radford2019language,peters2018deep,clark2019electra}. For a long time, supervised learning has been the most successful natural language learning paradigm. The pioneers of the pre-training idea challenged this view by showing that a vast amount of general knowledge about language, including both linguistic and commonsense knowledge, can be acquired by (pre-)training a model in a {\em task-agnostic} manner using {\em self-supervised} learning tasks. Self-supervised learning tasks are tasks for which the label associated with a training instance can be derived automatically from the text itself. Using Masked Language Model (MLM) and Next Sentence Prediction (NSP) as pre-training objectives, BERT~\cite{devlin2019bert} became the ground-breaker of the ``pre-training + fine-tuning'' paradigm. And many variants of it achieve notable performance in a wide range of aspects.

Inspired by the great success achieved by pre-trained NLP models, researchers start to apply pre-training techniques to the code-related domains, then obtain remarkable improvement of the model in many aspects, such as performance and robustness~\cite{feng2020codebert,guo2020graphcodebert,niu2022spt}. By re-pre-train a BERT model using source code data, Kanade et al.~\cite{kanade2020cubert} propose CuBERT for a more efficient source code representation. The effectiveness of CuBERT is demonstrated by fine-tuning on five classification tasks and a multi-headed pointer prediction task. Instead of using source code only, CodeBERT~\cite{feng2020codebert} further uses text data to make itself a bimodal pre-trained model for code and natural language. CodeBERT is based on RoBERTa~\cite{liu2019roberta}, a variant of BERT, and replaces NSP with Replaced Token Detection (RTD, a pre-training task proposed by ELECTRA~\cite{clark2019electra}). In order to utilize the code structure information, GraphCodeBERT~\cite{guo2020graphcodebert} adopts the data flow of source code and introduces two new pre-training tasks for learning representation of the code structure. T5-learning~\cite{mastropaolo2021studying} pre-trains a T5 on natural language and source code, surpassing the baselines of that dataset on four code-related datasets.

In this work, rather than directly using pre-training tasks already available in the NLP domain, we design pre-training tasks for code- and input-related features.

\section{Approach}
\label{section:approach}

This section introduces PassSum: the overall model architecture (Section~\ref{section:architecture}), the input representation (Section~\ref{section:input}), as well as the two ideas it employs, namely Path Embedding (Section~\ref{section:path_embedding}), self-supervised pre-training (Section~\ref{section:pre_train}).

\subsection{Model Architecture}
\label{section:architecture}

Architecturally, PassSum is essentially a multi-layer Transformer~\cite{vaswani2017attention}. As far as the parameter setting of the encoder and decoder is concerned, PassSum follows the ``base'' version of Transformer: (1) the number of layers (i.e., Transformer blocks) $L=6$, (2) the size of model $d=512$, (3) the dimension of feed forward $d_{ff}=2048$, (4) the number of self-attention heads $h=8$, and (4) the dropout rate $p_{dropout}=0.1$. The total number of parameters is 78M.

\subsection{Input Representation}
\label{section:input}

\begin{figure}[t]
  \centering
  \includegraphics[width=0.47\textwidth]{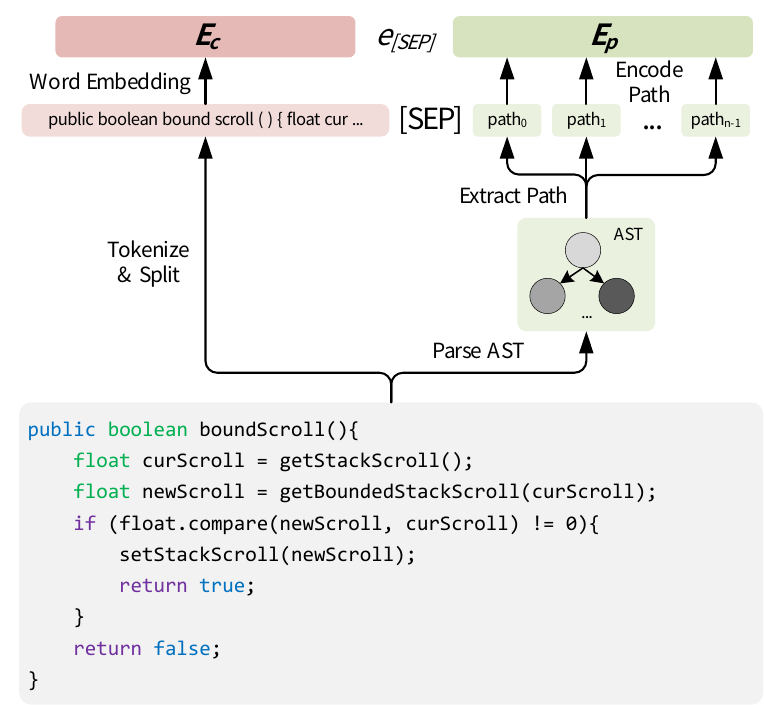}
  \caption{The input of a real world Java code snippet. Due to space constraints, we abbreviated the node names in the AST sequence. ``NL'' denotes the natural language input. We also do not show Positional Encoding layer.}
  \label{figure:model_input}
\end{figure}

The inputs of the model are two different types of components belonging to a complete \textit{Method} (i.e., it can be invoked by its name), namely, the code tokens, the path embedding vectors. In this subsection, we will demonstrate this with a real Java method\footnote{https://github.com/Unidata/thredds/blob/d2d68f9eee87f345625\\211324d71d5dc3e162ee1/cdm/src/main/java/thredds/client/catalog/\\Property.java\#L56-L63} shown in the bottom of \figurename~\ref{figure:model_input}.

As we can see from the figure, the first part of the input is the code token sequence of a method. We first split code tokens of the form \textit{CamelCase} and \textit{snake\_case} to respective subtokens. Finally, we input the token sequence $C=\{c_0,c_1,\dots,c_{m-1}\}$ into the word embedding layer to get the word embedding vector of the tokens, i.e., $\bm{E_c}\in \mathbb{R}^{m\times d}$, where $m$ is the number of code (sub)tokens. To distinguish the two parts of the input, a embedding vector of a special symbol ``[SEP]'', i.e., $\bm{e_{[SEP]}}\in \mathbb{R}^d$ is added to the end of $\bm{E_c}$.

The second part about the input will be covered in the next subsection. At last, we will get the encoded vectors $\bm{E_p}\in \mathbb{R}^{n\times d}$, where $n$ is the number of paths. And our final input will be represented as $Input=[\bm{E_c},\bm{e_{[SEP]}},\bm{E_p}]\in \mathbb{R}^{t\times d}$, where $t=m+1+n$.

\subsection{Path Embedding}
\label{section:path_embedding}

Although there are various ways to encode ASTs and good results have been achieved, there are still some problems with these methods.
\begin{enumerate}
    \item For methods that encode directly on the AST, the AST is usually very deep~\cite{lin2021improving}, which is rare in natural language syntax trees. This deep structure weakens the capability of tree-LSTMs to capture the complex semantics~\cite{zhu2015long}, leading to excessively slow training and loss of information.
    \item For methods that split an AST into several subtrees, they cannot capture the detailed relationship between the terminal nodes from different subtrees.
    \item For methods which use SBT to linearize ASTs, the sequences of SBT is too long, resulting in too slow training speed and high space occupation.
    \item For AST linearization methods like code2vec and code2seq, random sampling leads to unstable models.
\end{enumerate}

In view of the drawbacks mentioned above, we introduce an AST encoding approach called \textit{Path Embedding}, which aims to generate better embeddings for ASTs, as described below.

\subsubsection{Step 1: Extract Paths}
\label{section:extract_path}

\begin{figure*}[t]
    \centering
    \includegraphics[width=0.8\textwidth]{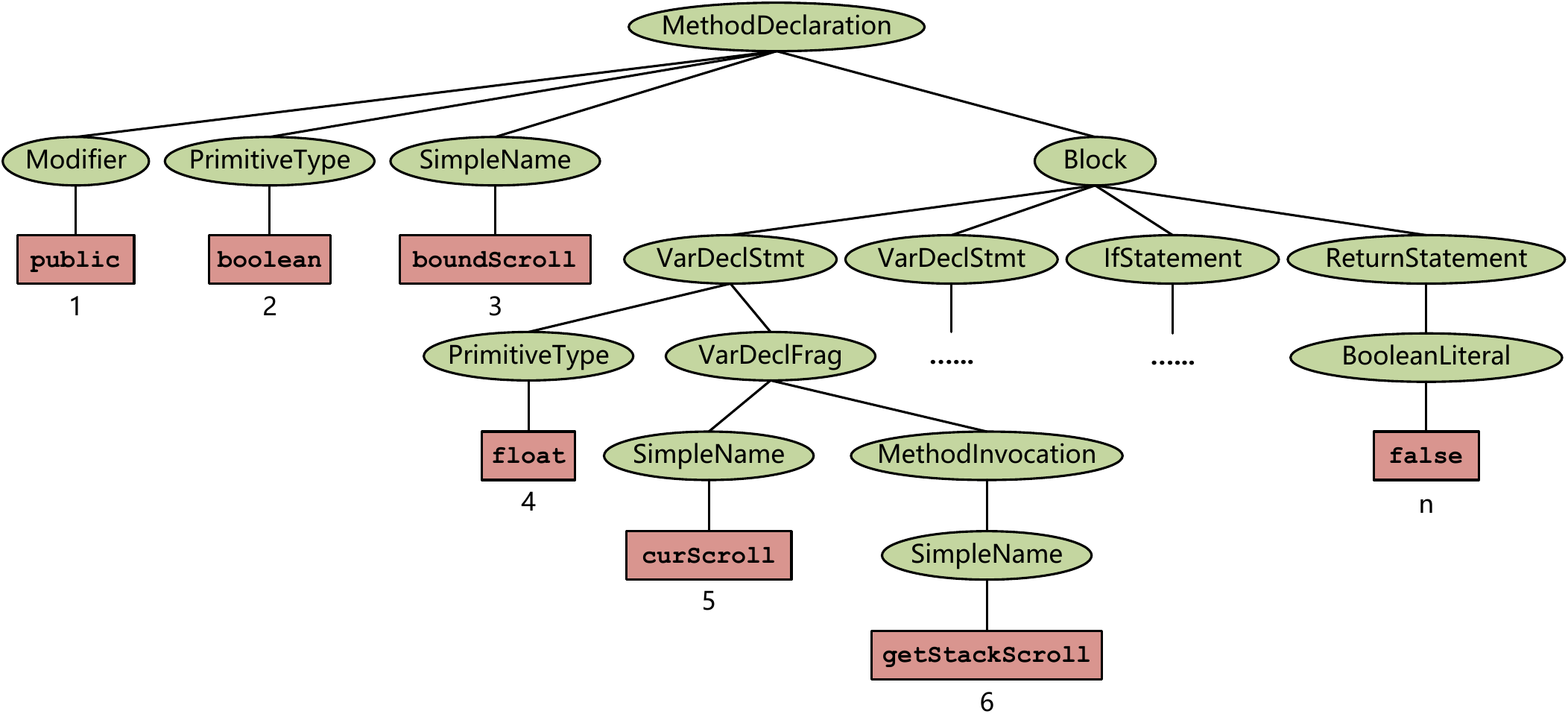}
    \caption{The AST of Java method {\tt boundScroll} in \figurename~\ref{figure:summarization_example}. Some node names use abbreviations, and subtrees of \textsf{VarDeclStmt} and \textsf{Ifstatement} are left out.}
    \label{figure:ast_example}
\end{figure*}

Take the previously mentioned Java method {\tt boundScroll} (see \figurename~\ref{figure:summarization_example}) as an example, the corresponding AST is shown in \figurename~\ref{figure:ast_example}. 

As we can see in the figure, the green ones are non-terminal nodes and the red ones are terminal nodes. We utilize acronyms for some nodes and exclude the subtrees of the second and third children of \textsf{Block} in the figure due to space restrictions. Assuming there are $n_t$ terminal nodes, we sort them from 1 to $n_t$ in the order of their corresponding identifiers appearing in the source code, and mark the number below the terminal nodes, denoting their position.

After converting the code to AST, we then perform the path extraction. A path $r$ is the path from the root node to a certain terminal node. Since the terminal node in the AST is important, inspired by code2vec, we divide a path into two parts, the first part is the path from the root node to the parent node of the terminal node, while the other part is the terminal node. As an example, the path to terminal node 5 in \figurename~\ref{figure:ast_example} is represented as:

\begin{center}
    \footnotesize ([\textsf{MethodDeclaraion}, \textsf{Block}, \textsf{VarDeclStmt}, \textsf{VarDeclFrag}, \textsf{SimpleName}], \texttt{curScroll})
\end{center}

One can see that each path is represented by a two-tuple, $r=(nodes,id)$. The first element is a list showing the path from the root node \textsf{MethodDeclaraion} to the parent node of the terminal node \textsf{SimpleName}, while the second element is the terminal node \texttt{curScroll}.

\begin{figure}[htbp]
    \centering
    \includegraphics[width=0.7\linewidth]{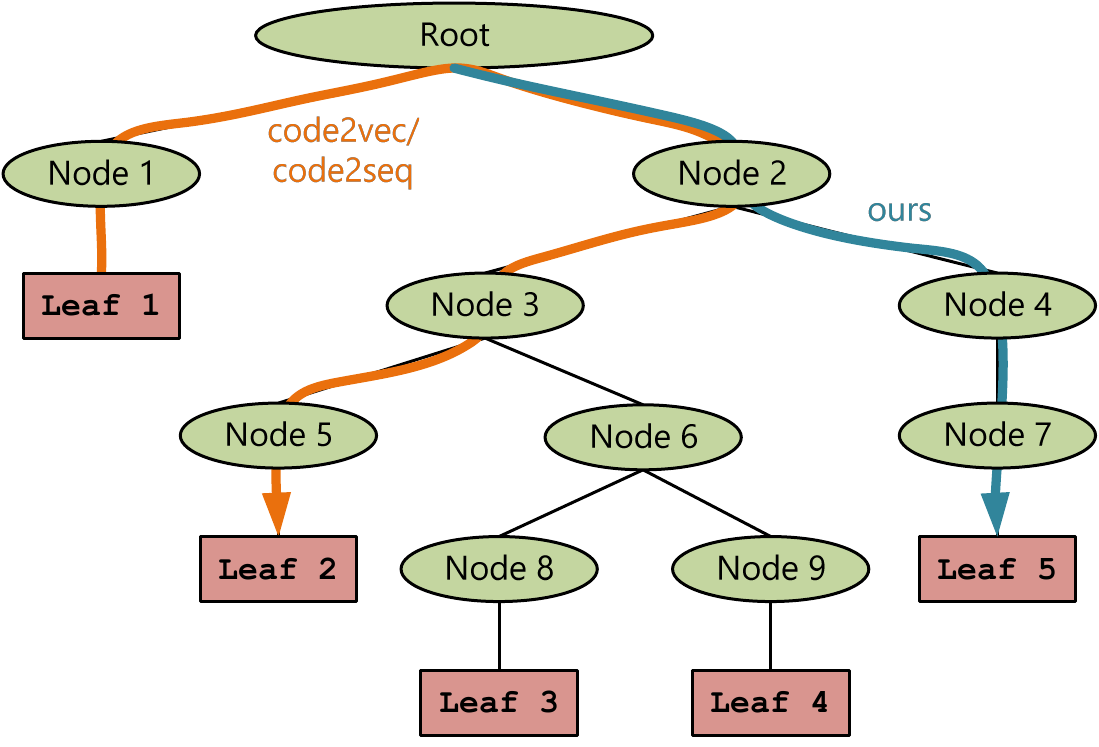}
    \caption{The comparison of code2vec/code2seq and our path extraction method.}
    \label{figure:path}
\end{figure}

\begin{figure}[htbp]
    \centering
    \includegraphics[width=0.8\linewidth]{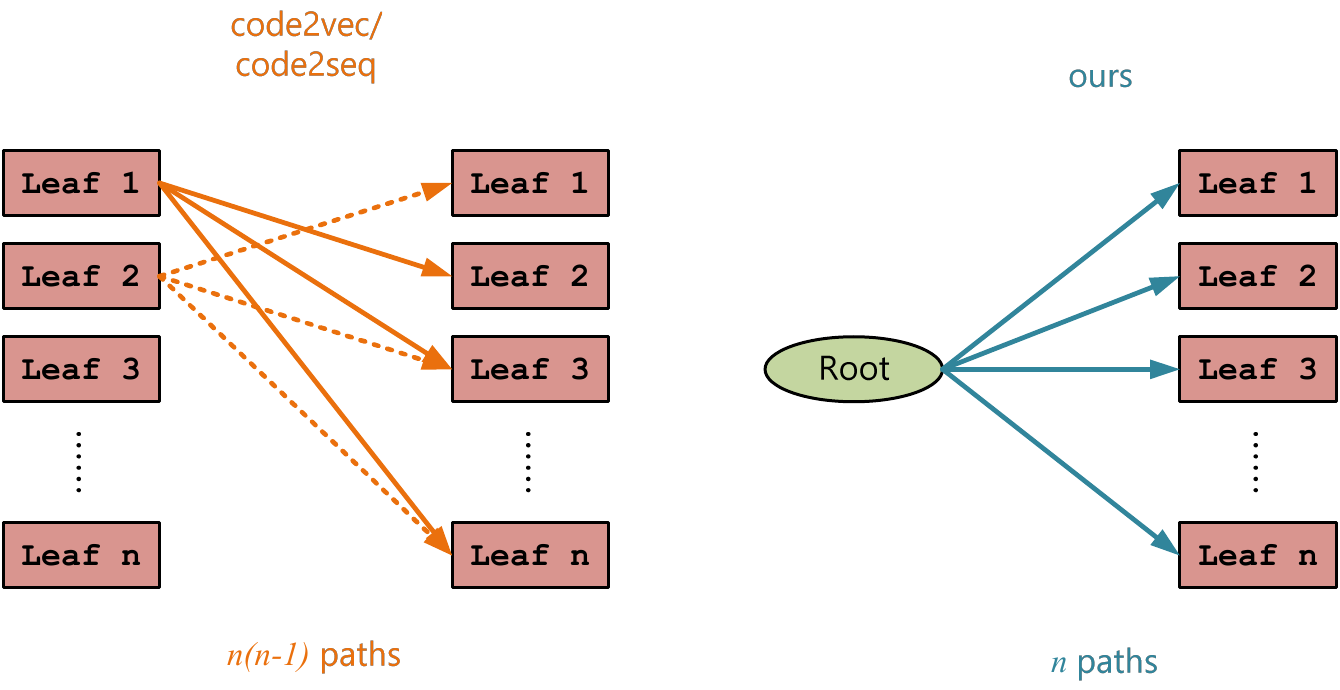}
    \caption{The comparison of the path extraction method of code2vec and ours.}
    \label{figure:path_comparison}
\end{figure}

One of advantages of our method of extracting paths compared to code2vec's method is that the number of extracted paths can be reduced from $P(n_t,2)$ (i.e., $n_t\cdot (n_t-1)$) to $n_t$. \figurename~\ref{figure:path} compares two path extraction methods. As for code2vec's method, it correlates any two different leaves and acts as a path. Instead, our method directly extract each path from the root to each leaf. As a result, if there are $n$ leaves in an AST, code2vec's method produces $P(n,2)=n(n-1)$ paths, while our method extracts only $n$ paths, as shown in \figurename~\ref{figure:path_comparison}.


Our method greatly reduces the size of the input compared to the path-context used by code2vec/code2seq. Without having to sample the input, the small size of the input increase the stability of the model. In addition, these paths are sorted by the position of the corresponding terminal nodes.

\subsubsection{Step 2: Embed Paths}
\label{section:embed_path}

\begin{figure}[t]
    \centering
    \includegraphics[width=0.9\linewidth]{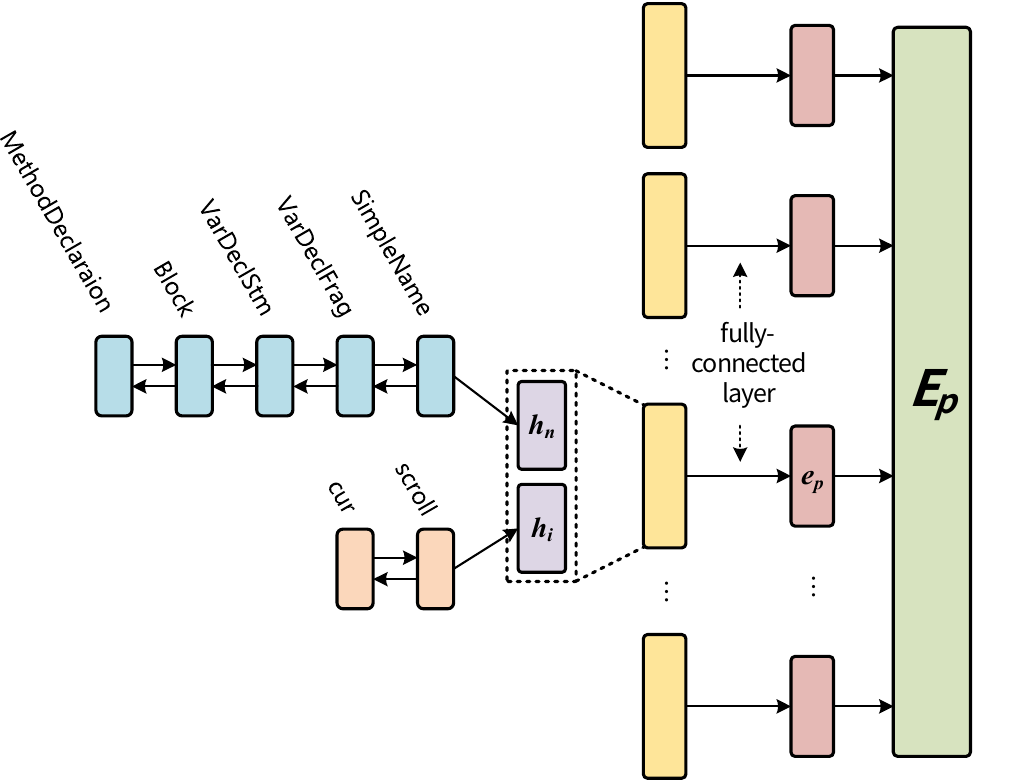}
    \caption{Illustration of Path Embedding. For a better presentation, we omit the embedding layer of two GRUs.}
    \label{figure:embed_path}
\end{figure}

An example of Path Embedding is illustrated in \figurename~\ref{figure:embed_path}.
For $n_t$ extracted paths, we use a code2seq-like approach to embed them. That is, we use two different GRUs to encode the two elements of the paths separately. For each $r=(nodes,id)$, $nodes=(node_1,node_2,\dots,node_{t_{node}})$ is a list of $t_{node}$ tokens so that we directly use a GRU to encode them into a vector $\bm{h_n}\in \mathbb{R}^d$. 
\begin{equation}
    \bm{h_n}=\textup{GRU}_\textup{n}(node_1,node_2,\dots,node_{t_{node}})
\end{equation}
Where $\textup{GRU}_\textup{n}$ is a bidirectional GRU encoder. 

While $id$ is an identifier, we firstly split it of the form \textit{CamelCase} and \textit{snake\_case} to respective subtokens, $id=(id_1,id_2,\dots,id_{t_{id}})$ where $t_{id}$ is the number of subtokens. Then this sequence of subtokens is encoded by another bidirectional GRU into a vector $h_i\in \mathbb{R}^d$.
\begin{equation}
    \bm{h_i}=\textup{GRU}_\textup{i}(id_1,id_2,\dots,id_{t_{id}})
\end{equation}
Splitting identifier can significantly reduce the size of vocabulary~\cite{hu2020deep,ahmad2020transformer}, allowing for a more accurate representation of identifiers.

After getting $h_n\in \mathbb{R}^d$ and $h_i\in \mathbb{R}^d$ for each path, we use a fully connected layer to project the concatenated two vectors into a new one.
\begin{equation}
    \bm{e_p}=\textup{sigmoid}(W_p[h_n,h_i]+b_p)
\end{equation}
Where $W_p\in \mathbb{R}^{2d\times d}$, $b_p\in \mathbb{R}^d$ are learnable parameters and $[h_n,h_i]\in \mathbb{R}^{2d}$ denotes the concatenation of $h_n$ and $h_i$. Then $\bm{e_p}$ is the embedded vector of a single path. As for $n_t$ paths, we will get a sequence of vectors, i.e., the matrix $E_p\in \mathbb{R}^{n_t\times d}$, each of which indicates each path obtained in Section~\ref{section:extract_path}. Then the vectors are passed through a positional encoding layer to inject positional information.

At a high level, Path Embedding takes as input a set of paths and returns an embedding vector for each of these paths. We then add the positional information to the embedding vector by passing them through the positional encoding layer. With this process, we incorporate the relative positional information of the terminal nodes, which indicates the position of the terminal nodes (i.e., identifiers) appearing in the code, into the embedded representations. These information is not considered by most current methods as they treat nodes at the same level as identical, unordered. Therefore, these information is not taken into account during the encoding, but such information is very important for program understanding. In addition, we use the self-attention mechanism in Transformer to automatically learn the relationship between paths, thus avoiding the key information that may be missing due to the random sampling.

\textcolor{red}{Please also note that when embedding all paths, the dependencies between variables will be unnoticed. For example, the ``define-use'' dependencies of a certain variable are not given explicitly in our input. But we believe that the self-attention mechanism in our model is able to automatically find valuable dependencies, since every representation vector of the path will imply information of the whole path. At the same time, we also believe that explicitly giving the dependencies between variables will further improve the performance of the model. We will leave it for subsequent work.}

\subsection{Pre-training Objectives}
\label{section:pre_train}

Recall that to make PassSum's performance more robust in the face of a small amount of training data, we pre-train PassSum using a large amount of unlabeled data composed solely of uncommented source code.

The question then is: how can we pre-train PassSum using unlabeled data? We do so by employing tasks that can be learned via \textit{self-supervised} learning. With self-supervised learning, we will generate labels from the input as pre-training targets to train the model to understand the input. For natural language, a very large number of studies have proposed various pre-training targets. These methods are designed to mine the input for knowledge that will enable the model to learn. In the following, we describe the three pre-training objectives we designed for our model, and for code summarization.

\subsubsection{Objective 1: Masked Language Model}
\label{section:pre_train_mlm}

\begin{figure}
    \centering
    \includegraphics[width=\linewidth]{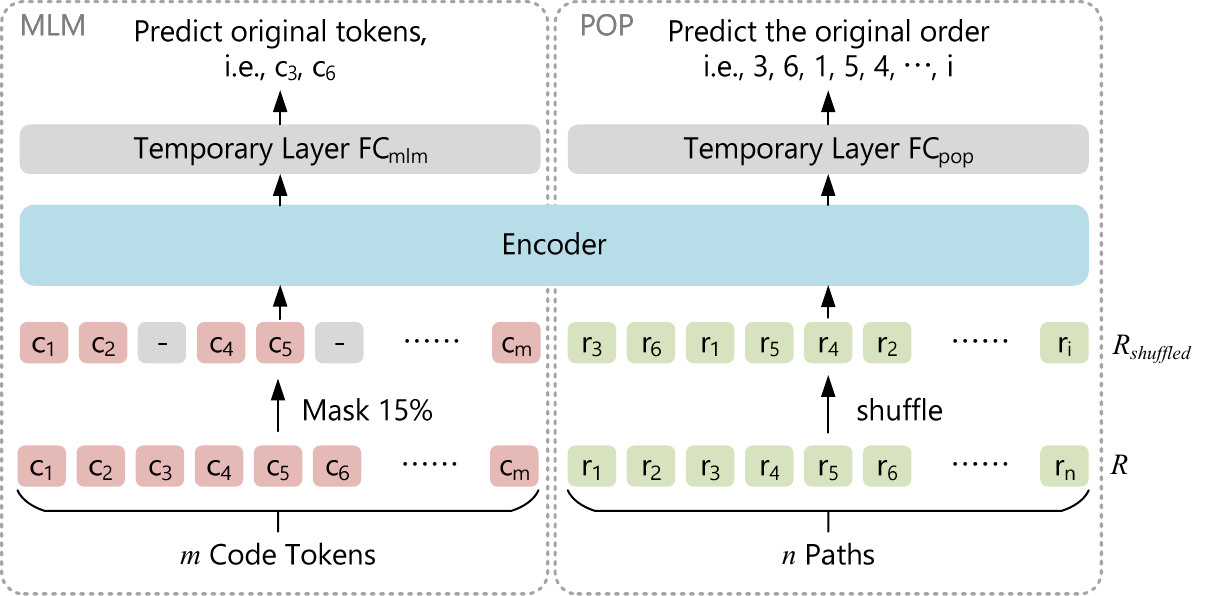}
    \caption{Application of MLM and POP pre-training objectives on encoder.}
    \label{figure:mlm_pop}
\end{figure}

Masked Language Model (MLM)~\cite{devlin2019bert}, arguably one of the most commonly used pre-training objectives, randomly masks a certain percentage of tokens in the input, making the model predict these masked tokens. The left side of \figurename~\ref{figure:mlm_pop} shows the process of MLM. We use MLM only in the first part of the input (i.e., code tokens) and follow the original settings. That is, 15\% of the tokens in the code token sequence are randomly masked, and the goal is to use the encoder predict the masked tokens. To enable encoder to output the probability of the vocabulary, we temporarily add a fully connected layer $\textup{FC}_\textup{mlm}$ after the last layer of encoder, so that the vocabulary probability
\begin{equation}
\begin{split}
    \bm{p_{mlm}}&=\textup{FC}_\textup{mlm}(\bm{O_{encoder}}\textup{[:m]})\\&=\bm{W_{mlm}}\bm{O_{encoder}}\textup{[:m]}+\bm{b_{mlm}}
\end{split}
\end{equation}
where $\bm{O_{encoder}}\in \mathbb{R}^{t\times d}$ is the output of the encoder, $\bm{O_{encoder}}\textup{[:m]}\in \mathbb{R}^{m\times d}$ denotes the first $m$ vectors in $\bm{O_{encoder}}$, i.e., the output corresponding to the first part of the input. $\bm{W_{mlm}}\in \mathbb{R}^{d\times |V|}$ and $\bm{b_{mlm}}\in \mathbb{R}^{|V|}$ are learnable parameters, where $V$ is the vocabulary, $|V|$ is the size of vocabulary. As a results, $\bm{P_{mlm}}\in \mathbb{R}^{m\times |V|}$. Since we only predict the tokens that are masked, so let $I_{mlm}$ be the set of indices of the masked tokens, the loss of MLM are calculated as
\begin{equation}
\begin{split}
    \mathcal{L}_\textup{mlm}=\sum_{i\in I_{mlm}}-\log \frac{\exp{(\bm{P_{mlm}}[\textup{i}][\textup{c}_\textup{i}])}}{\sum_{j=0}^{|V|-1} \exp{(\bm{P_{mlm}}[\textup{i}][\textup{j}])}}
\end{split}
\end{equation}

\subsubsection{Objective 2: Path Order Prediction}
\label{section:pre_train_pop}

Since each terminal node of the AST corresponds to an identifier in the code, there will be relative positions between paths, as we mentioned at the end of Section~\ref{section:path_embedding}. This order reflects the order in which each terminal node (corresponding to each path) appears in the code, so we shuffle the paths and let the model predict their original order, which is named as Path Order Prediction (POP) and shown in the right side of \figurename~\ref{figure:mlm_pop}.

Formally, the input list of $n$ path $R=\{r_0,r_1,\dots,r_{n-1}\}$ is shuffled to a new input list $R_{shuffled}=\{r_i,r_j,\dots,r_k\}$. We let 
$\textup{I}_{\textup{origin}}$ be the list of indices of each $r$ in $R_{shuffled}$ in the original $R$ after shuffling. 

Like in MLM, to enable encoder to output the probability of the original order, we temporarily add a fully connected layer $\textup{FC}_\textup{pop}$ after the last layer of encoder, so that the vocabulary probability
\begin{equation}
\begin{split}
    \bm{p_{pop}}&=\textup{FC}_\textup{pop}(\bm{O_{encoder}}\textup{[-n:]})\\&=\bm{W_{pop}}\bm{O_{encoder}}\textup{[-n:]}+\bm{b_{pop}}
\end{split}
\end{equation}
where $\bm{O_{encoder}}\textup{[-n:]}\in \mathbb{R}^{n\times d}$ denotes the last $n$ vectors in $\bm{O_{encoder}}$, i.e., the output corresponding to the second part of the input. $\bm{W_{pop}}\in \mathbb{R}^{d\times max\_path}$ and $\bm{b_{pop}}\in \mathbb{R}^{max\_path}$ are learnable parameters, where $max\_path$ is a hyperparameter which denotes the max number of paths in an input. As a results, $\bm{P_{mlm}}\in \mathbb{R}^{n\times max\_path}$. The loss of POP are calculated as
\begin{equation}
\begin{split}
    \mathcal{L}_\textup{pop}=\sum_{i=0}^{n-1}-\log \frac{\exp{(\bm{P_{pop}}[\textup{i}][\textup{I}_{\textup{origin}}[\textup{i}]])}}{\sum_{j=0}^{max\_path-1} \exp{(\bm{P_{pop}}[\textup{i}][\textup{j}])}}
\end{split}
\end{equation}

\subsubsection{Objective 3: Method Name Generation}
\label{section:pre_train_mng}

For each method written in programming language, there exists a natural label that does not exist in natural language, namely the method name. To the best of our knowledge, method names exist in any independently existing method (except anonymous methods/functions). Moreover, method names can be seen as concise versions of comments, both of which reflect the function and intent of the current method. Thus, making decoder to predict the method name allows the model to learn how to understand the function and intent of the method, so we let the model to generate the method name of a method whose method name is masked, i.e., Method Name Generation (MNG).

In addition, To enhance the sequence generation capability of the model, we split the method names into sequences of subtokens as well as the identifiers mentioned in Section~\ref{section:embed_path}. Hence we can think of the split method name as a short sequence, $Q={q_0,q_1,\dots,q_{s-1}}$, where $s$ is the length of this short sequence. We add special symbols ``[SOS]'' (start of sequence) and ``[EOS]'' (end of sequence) to the beginning and the end of $Q$ to get the input $Q_i$ and target $Q_t$ of the decoder, respectively.
\begin{equation}
\begin{split}
    Q_i=\{[\textup{SOS}],q_0,q_1,\dots,q_{s-1}\}\\
    Q_t=\{q_0,q_1,\dots,q_{s-1},[\textup{EOS}]\}
\end{split}
\end{equation}
where $Q_i$ is feed into the decoder, then the loss of MNG is calculated between the vocabulary distribution output by decoder $\bm{P_{mng}}\in \mathbb{R}^{(s+1)\times |V|}$ and $Q_t$:
\begin{equation}
\begin{split}
    \mathcal{L}_{mng}=\sum_{i=0}^{s+1}-\log \frac{\exp{(\bm{P_{mng}}[\textup{i}][Q_t[\textup{i}]]])}}{\sum_{j=0}^{s} \exp{(\bm{P_{mng}}[\textup{i}][\textup{j}])}}
\end{split}
\end{equation}

\begin{figure}
    \centering
    \includegraphics[width=0.8\linewidth]{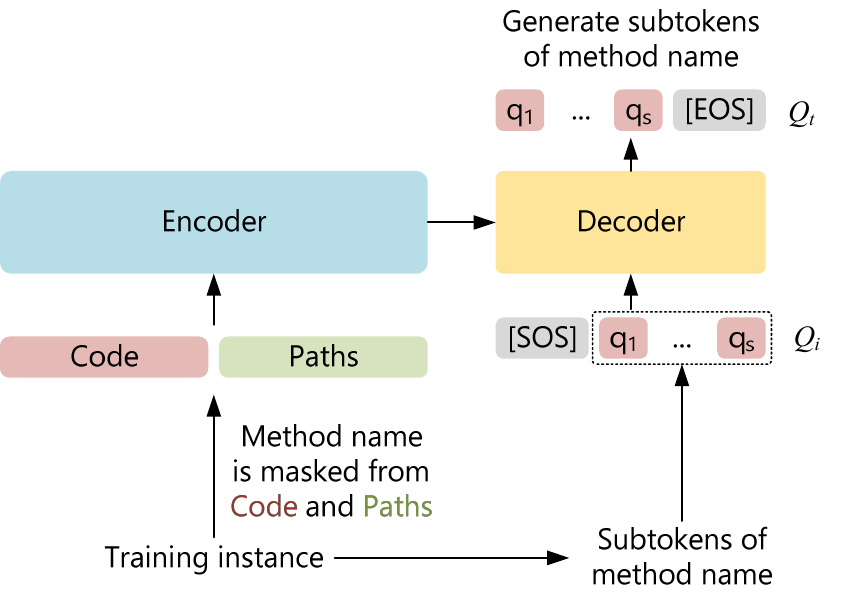}
    \caption{The MNG pre-training objective applied on the encoder-decoder framework.}
    \label{figure:mng}
\end{figure}

To better demonstrate, \figurename~\ref{figure:mng} illustrates the process of MNG.

Finally, when performing pre-training, we optimize these three losses with an equal probability, i.e., the total loss of pre-training is
\begin{equation}
    \mathcal{L}_{pre\_training}=\mathcal{L}_{mlm}+\mathcal{L}_{pop}+\mathcal{L}_{mng}
\end{equation}

\subsection{Fine-tuning}
\label{section:training}

By now we have constructed the complete model, and subsequently the model are pre-trained on the unlabeled dataset using the pre-training mentioned in Section~\ref{section:pre_train}. We then fine-tune the model on the labeled dataset with up to 200 epochs. The loss function is cross-entropy loss. Formally, like $\mathcal{L}_{mng}$ in Section~\ref{section:pre_train_mng}, when we have a code snippet and corresponding summary $y$ of length $q$, then we construct the input $y_i$ and target $y_t$ of decoder. When we feed the code snippet to the encoder, and the vocabulary distribution $\bm{P_{fine\_tuning}}\in \mathbb{R}^{(q+1)\times |V|}$ is output by the decoder, the loss is computed as:
\begin{equation}
    \mathcal{L}_{fine\_tuning}=\sum_{i=0}^{q+1}-\log \frac{\exp(\bm{P_{fine\_tuning}}[\textup{i}]y_t[\textup{i}])}{\sum_{j=0}^{q} \exp{(\bm{P_{fine\_tuning}}[\textup{i}][\textup{j}])}}
\end{equation}

\section{Evaluation}
\label{section:evaluation}

In this section, we evaluate PassSum, focusing on the following research questions:

\begin{itemize}
    \item RQ1: How effective is PassSum compared with the state-of-the-art baselines on the code summarization task?
    \item RQ2: Is there a performance advantage owing to the settings of the input representation used by PassSum?
    \item RQ3: Compared with other structural information encoding methods, does Path Embedding have advantages in time and space efficiency?
    \item RQ4: What are the impacts of pre-training on PassSum from performance, robustness, and convergence speed perspectives?
    \item RQ5: How effective is PassSum compared with the best baseline according to human evaluation?
\end{itemize}

\subsection{Experimental Setup}
\label{section:exp_setup}

\subsubsection{Dataset}
\label{section:dataset}

Our experiments are conducted on two popular programming languages, Java and Python. The labeled dataset and the unlabeled dataset are outlined as follows.

\textbf{Labeled Dataset.} Two public large-scale datasets in Java and Python are used as our labeled dataset. The Java dataset is provided by Hu et al.~\cite{hu2018summarizing} which contains Java methods and their comments from open source repositories in GitHub. The comment of each method is the first sentence extracted from its Javadoc. The Python dataset consists of the Python functions and comments collected by Barone et al.~\cite{miceli2017parallel}. We refer to these two datasets as JSCD (Java Code Summarization Dataset) and PCSD (Python Code Summarization Dataset), respectively

Besides, we also use CodeSearchNet~\cite{husain2019codesearchnet}, a widely used code summarization dataset. The CodeSearchNet corpus is programmatically obtained by scraping open-source repositories and pairing individual functions with their processed documentation as summary. It includes more than 6.4M codes from 6 programming language, i.e., Java, Python, JavaScript, PHP, Go and Ruby.

\textbf{Unlabeled Dataset.} The unlabeled Java dataset called \textit{Java-small} is provided by Alon et al.~\cite{alon2018code2seq}, which contains about 700K examples from the top-starred Java projects from GitHub. We use the training set as our unlabeled Java dataset. As for Python, Barone et al.~\cite{miceli2017parallel} provide not only parallel dataset but also monolingual dataset (namely code only). It is worth noting that these two datasets do not overlap, so we choose the monolingual dataset as our unlabeled Python dataset.

Table~\ref{table:dataset_statistics} shows the statistics of the labeled and unlabeled dataset.
\begin{table}[t]
    \centering
    \footnotesize
    \caption{Statistics of the labeled and unlabeled datasets.}
    \label{table:dataset_statistics}
    \begin{tabular}{llrrr}
        \toprule
                                       & Dataset    & Train   & Validation & Test   \\
        \midrule
            \multirow{8}{*}{Labeled}   & JCSD       & 69,708  & 8,714      & 8,714  \\
                                       & PCSD       & 106,271 & 1,947      & 1,944  \\
                                       & Java       & 164,923 & 5,183      & 10,955 \\
                                       & Python     & 251,820 & 13,914     & 14,918 \\
                                       & JavaScript & 58,025  & 3,885      & 3,291  \\
                                       & PHP        & 241,241 & 12,982     & 14,014 \\
                                       & Go         & 167,288 & 7,325      & 8,122  \\
                                       & Ruby       & 24,927  & 1,400      & 1,261  \\
        \midrule
            \multirow{2}{*}{Unlabeled} & Java       & \multicolumn{3}{r}{697,014}   \\
                                       & Python     & \multicolumn{3}{r}{125,566}   \\
        \bottomrule
    \end{tabular}
\end{table}

\subsubsection{Pre-processing}
\label{section:pre_processing}

We split source code and comment or name tokens to subtokens with the form \textit{CamelCase} and \textit{snake\_case}. This reduces the size of the vocabulary and increases the frequency of words in the vocabulary, thereby increasing the accuracy of the word embedding~\cite{gong2018frage}.

We use tree-sitter\footnote{https://tree-sitter.github.io/tree-sitter/} to get the AST of the code. We also discard a method if it satisfies one of the following conditions:

\begin{itemize}
    \item Constructor and main method for a class, because most of them are not functionally cohesive.
    \item Methods with annotations like @Test, @Bean, @Getter, @Setter and methods without these annotations but whose names start with ``test'', or start with ``get'', ``set'' and only have one line, because there is no need to generate comments for these methods.
    \item Methods with @Override annotation or the same name or comment (except digits) in a single file, we only keep the first instance to reduce redundancy.
\end{itemize}

We also conduct several processes to clean the comments. For example, we delete the tags and keep the content between them such as ``\textless code\textgreater'', ``\textless /code\textgreater '', ``\textless p\textgreater '', etc. Symbols like ``@see'', ``@link'' are deleted along with their contents.

For demonstration purpose, some detailed statistics of the JCSD and PSCD are shown in Table~\ref{table:ast_statistics}.
\begin{table}[t]
    \caption{Detailed statistics of the JCSD and PCSD datasets.}
    \label{table:ast_statistics}
    \centering
    \footnotesize
    \begin{tabular}{lll}
    \toprule
        Dataset  & JCSD         & PCSD       \\
    \midrule
        Avg. \# of terminal nodes & 119.50       & 54.83        \\
        Avg. depth of ASTs        & 9.50         & 8.63         \\
        Avg. length of paths      & 6.99         & 6.47         \\
    \hline
        Avg. length of code       & 120.10       & 169.29       \\
        Avg. length of labels     & 18.65        & 78.23        \\
    \bottomrule
    \end{tabular}
\end{table}

\subsubsection{Metrics}
\label{section:metrics}

We evaluate the source code summarization performance using three common automatic metrics, i.e., BLEU (B.)~\cite{papineni2002bleu}, METEOR (M.)~\cite{banerjee2005meteor}, and ROUGE-L (R.L)~\cite{lin2004rouge}. These metrics are also popular in machine translation, text summarization, and image captioning. We will also perform a human evaluation, the details of which are described in Section~\ref{section:rq5}.

\subsubsection{Baselines}
\label{section:baselines}

We use two types of baseline methods as our models for comparison in the evaluation.

The first type of methods are those without pre-training, which are trained directly on the labeled dataset, including:
\begin{itemize}
    \item \textbf{Code2seq}~\cite{alon2018code2seq} first extracts path-contexts from the AST, and for each path-context, averages the embedding vectors corresponding to the subtoken of the two terminal nodes to obtain the two vectors corresponding to two terminal nodes, then concatenates them with the vectors obtained from the LSTM encoded paths. Then the concatenation vectors are passed through the fully connected layer and the averaging operation to obtain the code representation vector. The summaries are then generated using a LSTM and Attention mechanism.
    \item \textbf{Rencos}~\cite{zhang2020retrieval} combines IR and neural network methods. Firstly, it trains an attentional seq2seq model. During testing, for each sample, Rencos retrieves its two most similar code snippets in the training set from the aspects of syntax and semantics respectively, then encodes these three code snippets and predicts the summary by fusing them during decoding.
    \item \textbf{Code+gnn+GRU}\cite{leclair2020improved} uses GRUs and ConvGNNs to encode the code sequence and AST respectively.
    \item \textbf{NeuralCodeSum}~\cite{ahmad2020transformer} employs a Transformer-based model by incorporating relative position representations and copy attention~\cite{see2017get}.
    \item \textbf{SiT}~\cite{wu2021code} presents a novel structured-induced Transformer model on code summarization task. The model effectively incorporate multi-view structure into attention mechanism, and aggregate both global self-attention and structure-induced self-attention representations.
\end{itemize}

Another type of baselines are pre-trained models we introduced in Section~\ref{section:bg_pre_train}, which are first pre-trained on unlabeled datasets and then fine-tuned on labeled datasets, including \textbf{CodeBERT}~\cite{feng2020codebert}, \textbf{GraphCodeBERT}~\cite{guo2020graphcodebert} and \textbf{T5-learning}~\cite{mastropaolo2021studying}.

We use the official source code provided by authors for experiments. To ensure fairness of the comparison, we keep all baseline models using the dataset we introduce in Section~\ref{section:dataset} for pre-training (only for the second type of baseline) as well as for fine-tuning the dataset. Since each model uses different neural networks, we keep their model dimensions as 512. This will result in approximately the same number of parameters for all Transformer-based models, i.e., models except for code2seq, ASTNN, Rencos and code+gnn+GRU. While these four models have less parameters.

\subsubsection{Settings}
\label{section:settings}

We implement our model using PyTorch 1.7.1\footnote{https://pytorch.org/}. We set the maximum length of code and comments to 200 and 30 respectively. The maximum number of paths for one sample is 100. In the case of each path $r=(nodes,id)$, the maximum lengths of $nodes$ and $id$ (after splitting) are 30 and 10 respectively. The dimensions of embedding, Transformer, $d$ in Section~\ref{section:embed_path} are all set to 512. The dimension of the Feed Forward sublayer is set to 2048. The batch size is set to 32 and the beam size is 5 when testing. Vocabulary of methods without pre-training is built upon the training set, and for methods require pre-training, the vocabulary is built on the pre-training set. The training phase after pre-training directly uses the source and target vocabularies constructed for pre-training. And the sizes of vocabulary are all set to 30,000.

The optimizer of both pre-training and fine-tuning is Adam~\cite{kingma2017adam} with default parameters. The first 2000 steps are used to warm up, during which the learning rate increases linearly from 0 to the initial learning rate. For pre-training, the number of epochs is set to 50. And for fine-tuning, the maximum epoch is set to 200, we also adopt early stopping if the validation performance does not increase for 20 consecutive epochs. All experiments are performed on 4$\times$NIVDIA V100 32Gs with a total batch size of 256 and with mixed precision.

\subsection{Results}
\label{section:results}

In this subsection, we answer our research questions.

\subsubsection{RQ1: How effective is PassSum compared with the state-of-the-art baselines on the code summarization task?}
\label{section:rq1}

Experimental results expressed in terms of the three automatic evaluation metrics, are shown in Table~\ref{table:exp_results} and Table~\ref{table:exp_csn_results}. 

\begin{table}[t]
    \caption{Results of baselines and PassSum for code summarization. Due to the data pre-processing problem, we could not perform the experiments of ASTNN and code+gnn+GRU on the Python dataset. PassSum outperforms baselines significantly ($p<0.05$).}
    \label{table:exp_results}
    \centering
    % \footnotesize
    \resizebox{\linewidth}{!}{%
    \begin{tabular}{lcccccc}
        \toprule
            \multicolumn{1}{l}{\multirow{2}{*}{Methods}} &
            \multicolumn{3}{c}{Java} &
            \multicolumn{3}{c}{Python} \\ \cline{2-7}
            \multicolumn{1}{c}{} &
            \multicolumn{1}{c}{B.} &
            \multicolumn{1}{c}{M.} &
            \multicolumn{1}{c}{R.L} &
            \multicolumn{1}{c}{B.} &
            \multicolumn{1}{c}{M.} &
            \multicolumn{1}{c}{R.L} \\
        \midrule
            code2seq            & 24.42 & 15.35 & 33.95 & 17.54 & 08.49 & 20.93 \\
            ASTNN               & 25.40 & 16.92 & 36.10 & -     & -     & -     \\
            Rencos              & 36.95 & 26.51 & 51.22 & 28.91 & 20.47 & 45.00 \\
            code+gnn+GRU        & 30.55 & 18.95 & 42.65 & -     & -     & -     \\
            NeuralCodeSum       & 44.58 & 26.43 & 54.76 & 32.52 & 19.77 & 46.73 \\
            SiT                 & 45.76 & 26.58 & 55.58 & 34.11 & 21.11 & 48.35 \\
        \hline
            CodeBERT            & 45.33 & 26.20 & 54.61 & 33.46 & 21.65 & 49.36 \\
            GraphCodeBERT       & 46.03 & 26.45 & 56.51 & 33.89 & 22.12 & 50.44 \\
            T5-learning         & 43.74 & 26.29 & 53.64 & 32.20 & 21.99 & 49.31 \\
        \hline
            PassSum             & \textbf{48.01} & \textbf{29.61} & \textbf{58.23} & \textbf{35.14} & \textbf{24.90} & \textbf{51.76} \\
            ~- w/o pre-training & 46.15 & 27.59 & 56.00 & 33.48 & 23.14 & 49.28 \\
        \bottomrule
    \end{tabular}
    }
\end{table}

\begin{table*}[t]
    \centering
    \caption{Results of baselines and PassSum on the CodeSearchNet dataset. Due to the data pre-processing problem, we only conduct the experiment on code2seq for the first baseline group. PassSum outperforms baselines significantly ($p<0.05$).}
    \label{table:exp_csn_results}
    \resizebox{\linewidth}{!}{%
    \begin{tabular}{lrrrrrrrrrrrrrrrrrr}
        \toprule
            \multirow{2}{*}{Methods} &
            \multicolumn{3}{c}{Java} &
            \multicolumn{3}{c}{Python} &
            \multicolumn{3}{c}{JavaScript} &
            \multicolumn{3}{c}{PHP} &
            \multicolumn{3}{c}{Go} &
            \multicolumn{3}{c}{Ruby} \\
        \cline{2-19}
                              & B.   & M.   & R.L  & B.   & M.  & R.L  & B.   & M.   & R.L  & B.   & M.   & R.L  & B.   & M.   & R.L  & B.  & M.  & R.L  \\
        \midrule
            code2seq          & 12.8 & 15.0 & 30.9 & 9.4  & 5.2 & 20.1 & 9.4  & 8.9  & 21.1 & 19.4 & 18.1 & 39.0 & 14.8 & 10.2 & 29.0 & 7.5 & 6.9 & 15.8 \\
            CodeBERT          & 13.6 & 16.4 & 32.2 & 10.7 & 6.0 & 21.0 & 10.0 & 10.5 & 22.2 & 20.1 & 19.3 & 42.7 & 16.0 & 11.7 & 30.4 & 8.3 & 7.5 & 17.2 \\
            GraphCodeBERT     & 14.5 & 17.8 & 33.5 & 11.0 & 7.4 & 22.1 & 10.9 & 11.4 & 24.9 & 20.0 & 20.2 & 42.8 & 16.5 & 12.5 & 30.1 & 8.3 & 8.1 & 18.0 \\
            T5-learning       & 13.5 & 16.8 & 34.2 & 9.7  & 5.5 & 16.3 & 9.0  & 9.2  & 19.6 & 18.4 & 17.9 & 35.6 & 15.3 & 11.0 & 25.9 & 7.6 & 6.4 & 15.1 \\
        \hline
            PassSum           & \textbf{15.8} & \textbf{18.9} & \textbf{34.8} & \textbf{12.0} & \textbf{8.0} & \textbf{22.9} & \textbf{13.0} & \textbf{13.3} & \textbf{26.8} & \textbf{22.1} & \textbf{21.8} & \textbf{44.6} & \textbf{19.0} & \textbf{13.8} & \textbf{31.8} & \textbf{9.0} & \textbf{9.5} & \textbf{19.0} \\
            ~-w/o pre-training & 14.8 & 18.0 & 33.7 & 11.4 & 7.5 & 22.5 & 11.1 & 11.8 & 25.0 & 20.5 & 20.3 & 43.7 & 16.8 & 12.6 & 30.2 & 7.4 & 7.5 & 17.1 \\
        \bottomrule
    \end{tabular}
    }
\end{table*}

First of all, fine-tuned GraphCodeBERT outperforms all baselines. We can also see that PassSum outperforms state-of-the-art competitors, which we attribute to the path embedding and pre-training approaches we employ. PassSum learns the representation of code sequences and ASTs from a vast quantity of data during the pre-training process, as well as the relationship between the code and its structural information.

Furthermore, we observe that PassSum improves more on the JCSD (2.24) than it does on Python (1.71). We believe this is due to the beneficial condition of Path Embedding that we employ.

We know from Section~\ref{section:path_embedding} that Path Embedding can shorten the length of the input, and we find that the average number of terminal nodes of the AST for JCSD (119.50) is much higher than that of the PCSD AST (54.83) (See Table~\ref{table:ast_statistics}). This indicates shows that the size of ASTs in Java datasets is generally larger than that of Python ASTs, leading to more benefits of Path Embedding on the Java dataset, and thus more improvement of PassSum.

Table~\ref{table:dataset_statistics} also reveals a possible factor, which is the magnitude of the unlabeled data. The Python unlabeled dataset is substantially smaller than the Java unlabeled dataset used for pre-training. PassSum's knowledge learned on Python through pre-training may be less than that learned on Java, resulting in a smaller improvement of PassSum on Python than on Java. Table~\ref{table:exp_csn_results} leads to similar conclusions as well.

Furthermore, we can see from Table~\ref{table:exp_csn_results} that PassSum performs better on languages not seen in the pre-training phase, namely JavaScript, PHP, Go, and Ruby. This demonstrates PassSum's ability to generalize between languages. This, we believe, is attributable to two factors. Firstly, our input representation, AST, is our main input portion, and ASTs do not change greatly between languages, allowing PassSum to conceal programming language differences to some extent. The second is pre-training, which provides PassSum with a general understanding of programming languages so that it can migrate to programming languages it has never encountered before.

As for the effectiveness of pre-training and Path Embedding, we will evaluate and analyze them in the next two research questions.

\subsubsection{RQ2: Is there a performance advantage owing to the settings of the input representation used by PassSum?}
\label{section:rq2}

We conduct experiments in two ways to further investigate the effectiveness of the input representation of PassSum: first, we evaluate the effectiveness of Path Embedding and its superiority over other methods; second, we compare the performance of the concatenation of two inputs we used, and the performance of using two encoders to process the two inputs separately. Note that in this research problem, all methods do not perform pre-training. Results are shown in Table~\ref{table:exp_input}.
\begin{table}[t]
    \caption{Ablation study on the input representation of PassSum.}
    \label{table:exp_input}
    \centering
    \footnotesize
    \begin{tabular}{lcccccc}
        \toprule
            \multicolumn{1}{l}{\multirow{2}{*}{Methods}} &
            \multicolumn{3}{c}{Java} &
            \multicolumn{3}{c}{Python} \\ \cline{2-7}
            \multicolumn{1}{c}{} &
            \multicolumn{1}{c}{B.} &
            \multicolumn{1}{c}{M.} &
            \multicolumn{1}{c}{R.L} &
            \multicolumn{1}{c}{B.} &
            \multicolumn{1}{c}{M.} &
            \multicolumn{1}{c}{R.L} \\
        \midrule
            PassSum       & 46.15 & 27.59 & 56.00 & 33.48 & 23.14 & 49.28 \\
        \hline
            \multicolumn{7}{l}{Path Embedding is replaced by} \\
        \hline
            ~- None           & 41.21 & 24.46 & 53.09 & 31.14 & 19.60 & 47.54 \\
            ~- SBT            & 42.82 & 24.45 & 53.73 & 31.45 & 20.49 & 48.32 \\
            ~- SBT (600)      & 42.88 & 24.47 & 53.98 & 31.50 & 20.48 & 48.55 \\
            ~- code2seq       & 45.24 & 26.87 & 55.02 & 32.53 & 21.45 & 48.85 \\
            ~- code2seq (200) & 45.40 & 26.89 & 55.27 & 32.62 & 21.60 & 49.05 \\
        \hline
            \multicolumn{7}{l}{multi-encoder~\cite{libovicky2018input}} \\
        \hline
            ~- serial       & 44.77 & 26.18 & 54.26 & 32.02 & 20.35 & 48.04 \\
            ~- parallel     & 44.72 & 26.14 & 54.23 & 31.99 & 20.30 & 48.06 \\
            ~- flat         & 44.06 & 25.80 & 53.83 & 31.57 & 20.05 & 47.69 \\
            ~- hierarchical & 44.68 & 26.11 & 54.16 & 32.03 & 20.39 & 48.06 \\
        \bottomrule
    \end{tabular}
\end{table}

First of all, ``PassSum None'' indicates that we delete the Path Embedding from the PassSum input, leaving only the first portion of the input, the code tokens. There is a considerable fall in model performance after eliminating path embedding, indicating that our second input, Path Embedding, can help the model grasp the structural information of the code and generate better summaries.

The Path Embedding is then replaced by SBT~\cite{hu2020deep} and code2seq~\cite{alon2018code2seq}. For a fair comparison, we keep the maximum length of the second half of the inputs at 100, as stated in Section~\ref{section:settings}. However, we find that the length of the SBT far surpasses 100 in most situations, and that the length of the SBT is more than three times the length of the code tokens on average. As a result, we set the maximum length of SBT to 600 and refer to the results as ``PassSum SBT (600)''. However, based on the results, the improvement produced by increasing the maximum length of the SBT is rather minor. We contend that, while the increased SBT length enriches the input with more structural information, the excessive length exploits the difficulty of the model in extracting functional features.

Furthermore, even after replacing Path Embedding with code2seq, we notice a minor loss in performance. One probable explanation is that the code2seq input length is more than the limit we set, i.e., 100, which may result in some critical information in the AST being cropped out. As a result, we increased the input length to 200, which the authors of code2seq determined to be a good sweet spot between capturing enough information and making training practical in the GPU's memory. As seen in the table, ``PassSum code2seq (200)'' shows a minor performance gain over ``PassSum code2seq'', which we assume is due to the longer input length increasing the probability that relevant information may be fed into the model.

Next, we separate the two parts of the input, use two Transformer Encoders, and experiment with the four combination strategies mentioned in~\cite{libovicky2018input}. As can be seen, ``flat'' performs somewhat worse than the other three approaches, while the other three strategies perform around the same. At the same time, the performance of all four solutions is inferior to that of PassSum's concatenation method. The reason we think lies in the difference between using a single encoder and multi-encoders. A single encoder conducts the interactions between feature extractors for extracting relevant features from two types of inputs, namely code tokens (lexical features) and linearized AST (syntactic features), considerably earlier and more comprehensively than multi-encoders. In other words, if different encoders are used for different inputs, the two inputs will not interact in time, resulting in a diminished mutual influence while extracting features.

\subsubsection{RQ3: Compared with other encoding methods for structural information, does Path Embedding have advantages in time and space efficiency?}%可以吗？
\label{section:rq3}

We analyzed the shortcomings of other AST encoding methods in Section~\ref{section:path_embedding}, which is why we use Path Embedding, and we can learn the advantages of Path Embedding in terms of performance by comparing RQ1 and RQ2. In this research question, we will examine whether Path Embedding still has advantages in time and space.

First, the models that we selected for comparison are ASTNN, code+gnn+GRU, PassSum SBT, PassSum SBT (600) and PassSum code2seq. We run all the models and PassSum on a single NVIDIA V100 with a batch size of 64. We run all models for 10 epochs and then calculate their average training speed, as well as the average GPU memory usage. The results are shown in Table~\ref{table:exp_time_space}.
\begin{table}[t]
    \caption{Training speed and GPU memory usage of models.}
    \label{table:exp_time_space}
    \centering
    \footnotesize
    \begin{tabular}{lrr|rr}
        \toprule
            \multirow{2}{*}{Methods} &
            \multicolumn{2}{c|}{\begin{tabular}[c]{@{}c@{}}Training Speed\\(s/epoch)\end{tabular}} &
            \multicolumn{2}{c}{\begin{tabular}[c]{@{}c@{}}GPU Memory Usage\\(MB)\end{tabular}} \\ \cline{2-5} &
            \multicolumn{1}{c}{Java} &
            \multicolumn{1}{c|}{Python} &
            \multicolumn{1}{c}{Java} &
            \multicolumn{1}{c}{Python} \\
        \midrule
            ASTNN                  &   949 &     - & 21,174 &      - \\
            code+gnn+GRU           & 1,015 &     - & 27,315 &      - \\
            PassSum SBT            & 1,094 &   768 & 21,894 & 19,046  \\
            PassSum SBT (600)      & 1,943 & 1,278 & 31,040 & 28,187  \\
            PassSum code2seq       & 2,482 & 1,615 & 25,176 & 21,406  \\
            PassSum code2seq (200) & 2,510 & 1,629 & 27,453 & 23,774  \\
            PassSum                & 1,135 &   791 & 22,798 & 19,973  \\
        \bottomrule
    \end{tabular}
\end{table}

First, we can observe that ``PassSum SBT'' is the most time and space efficient, because SBT is extracted from AST prior to training and may be directly input into Transformer without going through any neural network. Furthermore, when utilizing SBT, we do not need to include any extra neural networks outside of Transformer, such as Path Embedding or code2seq. However, as we know from RQ2, using SBT results in a considerable loss in performance, even when its maximum length is increased to 600, and the time and space occupation increases dramatically.

PassSum, with the exception of ``PassSum SBT'', has the highest time and space efficiency. As neural networks that can extract features directly from AST, ASTNN and code+gnn+GRU are significantly less time and space efficient, confirming the conclusion made in Section~\ref{section:path_embedding}. Finally, ``PassSum code2seq'' is inferior to PassSum because to the excessive amount of path-contexts, as well as the fact that the path-context is often longer than the path utilized for Path Embedding, resulting in too slow LSTM propagation.

Finally, when viewed as a whole, the model is more efficient in time and space on Python than it is on Java. We believe this is because, as shown statistically in Table~\ref{table:ast_statistics}, the AST of Python is on average smaller than that of Java, in terms of the number of terminal nodes, the depth of the AST, and the length of the path.

\subsubsection{RQ4: What are the impacts of pre-training on PassSum from \textbf{Performance}, \textbf{Robustness}, and \textbf{Convergence Speed} perspectives?}
\label{section:rq4}

In this RQ, we discuss several aspects of pre-training that have an impact on PassSum, including performance, robustness, and speed of convergence on the code summarization task.

\textit{Performance.} First of all, the ``PassSum'' and ``PassSum w/o pre-training'' rows of Table~\ref{table:exp_results} show the result of PassSum \textit{with} and \textit{without} pre-training. As we can see, the scores of all metrics improve a lot with pre-training, which illustrates that pre-training can indeed improve PassSum's performance.

\begin{figure*}[htbp]
    \centering
    \subfloat[Java code summarization.]{
        \includegraphics[width=0.32\textwidth]{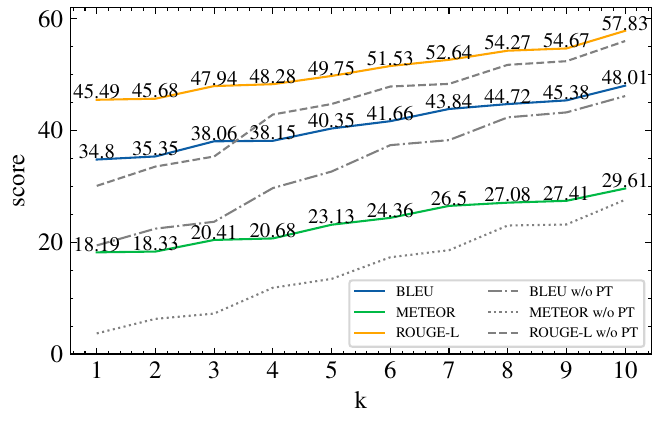}
        \label{figure:java_subset}
    }
    \subfloat[Python code summarization]{
        \includegraphics[width=0.32\textwidth]{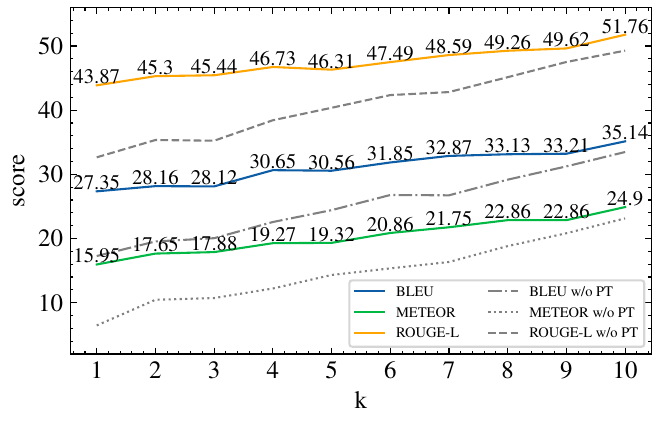}
        \label{figure:python_subset}
    }
    \subfloat[Ruby code summarization]{
        \includegraphics[width=0.35\textwidth]{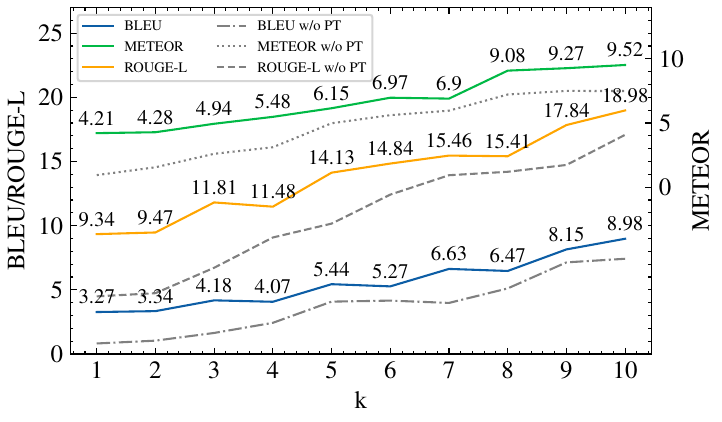}
        \label{figure:ruby_subset}
    }
    \caption{Results when $k$ goes from 1 to 10. The solid line is the result of the pre-trained PassSum, the dashed line is the result of PassSum without pre-training.}
    \label{figure:exp_subset}
\end{figure*}

\textit{Robustness.} Recall that one goal of self-supervised pre-training is to solve downstream tasks with high performance even if there is only little labeled data available for training a dedicated model. So, how well can a model withstand a lack of training data while still performing well on downstream tasks is a crucial consideration when evaluating pre-training models. As a result, we narrow down the training data for code summarization and observe how the results change. We train on $k/10, (k=1,2,...,9)$ data from the training set each time, then test on the same whole testing set. We run experiments on two models, one with a pre-trained PassSum and the other without. The results are shown in \figurename~\ref{figure:java_subset} and \figurename~\ref{figure:python_subset}.

Except for a few situations, it is clear that the performance of both models shows a diminishing trend in all evaluation metrics when $k$ declines (i.e., the size of the training set diminishes). It demonstrates that, even for pre-trained models, the size of the training set has an impact on performance. When we compare the magnitudes of the two models' declines, we can observe that the gap in their performance grows as $k$ lowers. We may conclude that, while the pre-trained PassSum result lowers, pre-training can maintain the robustness of PassSum to some extent when the fine-tuned dataset is reduced.

However, the above smaller training datasets are artificial. In addition, Java and Python are still large programming languages, and the pre-training dataset we use is also in these two languages. Therefore, we use another small real-world programming language for our experiments, which is Ruby. The dataset is the Ruby dataset in CodeSearchNet, which contains 24,927 training samples, 1,400 validation samples and 1,261 testing samples. The results of baselines and PassSum on Ruby dataset can be find in the right side of Table~\ref{table:exp_csn_results}.
% \begin{table}[t]
%     \caption{Results on CodeSearchNet Ruby dataset.}
%     \label{table:exp_ruby}
%     \centering
%     \footnotesize
%     \begin{tabular}{lcccccc}
%         \toprule
%             \multicolumn{1}{l}{\multirow{2}{*}{Methods}} &
%             \multicolumn{3}{c}{Ruby} \\ \cline{2-4}
%             \multicolumn{1}{c}{} &
%             \multicolumn{1}{c}{B.} &
%             \multicolumn{1}{c}{M.} &
%             \multicolumn{1}{c}{R.L} \\
%         \midrule
%             CodeBERT            & 8.25 & 7.53 & 17.24 \\
%             GraphCodeBERT       & 8.34 & 8.14 & 18.01 \\
%             T5-learning         & 7.58 & 6.36 & 15.14 \\
%         \hline
%             PassSum             & 8.98 & 9.52 & 18.98 \\
%             ~- w/o pre-training & 7.42 & 7.49 & 17.08 \\
%         \bottomrule
%     \end{tabular}
% \end{table}

We also conduct experiments on the relationship between ruby dataset size and final results, and present the results in \figurename~\ref{figure:ruby_subset}.
% \begin{figure}
%     \centering
%     \includegraphics[width=0.45\textwidth]{figures/ruby_subset.pdf}
%     \caption{Results on Ruby code summarization when $k$ goes from 1 to 10. The solid line is the result of the pre-trained PassSum, the dashed line is the result of PassSum without pre-training}
%     \label{figure:ruby_subset}
% \end{figure}

It can be seen that PassSum still outperforms other pre-training baseline models on a small real-world programming language dataset. Furthermore, it is shown that pre-training can still improve the robustness of PassSum in this scenario. \figurename~\ref{figure:ruby_subset} also shows that models on the Ruby dataset, as well as the java and python datasets, exhibit comparable connections between training and size and end results. Only the trend of the results on ruby fluctuates more.

\begin{figure*}[t]
    \centering
    \subfloat[Java code summarization.]{
        \includegraphics[width=0.33\textwidth]{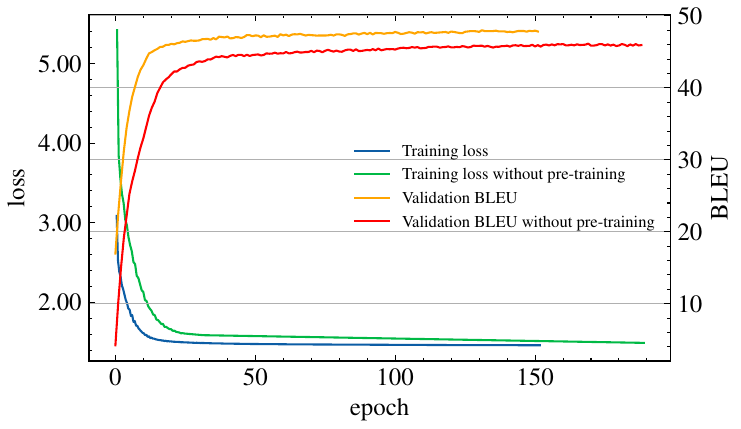}
        \label{figure:java_curves}
    }
    \subfloat[Python code summarization]{
        \includegraphics[width=0.33\textwidth]{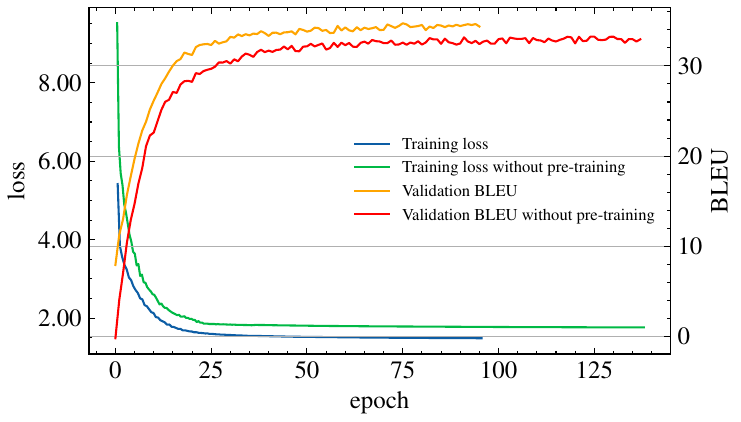}
        \label{figure:python_curves}
    }
    \subfloat[Ruby code summarization]{
        \includegraphics[width=0.33\textwidth]{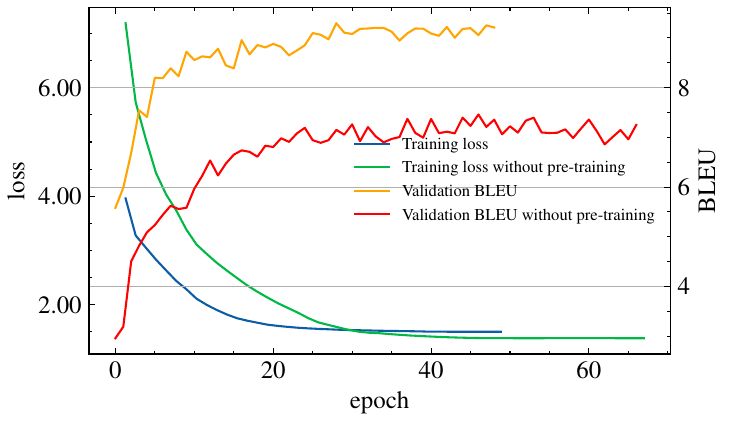}
        \label{figure:ruby_curves}
    }
    \caption{Learning curves of the pre-trained PassSum and the randomly initialized PassSum on the code summarization.}
    \label{figure:loss_curves}
\end{figure*}

\textit{Convergence Speed.} There will still be those who doubt the utility of pre-training for PassSum, because we have only demonstrated that pre-training aids PassSum in the final result, while we have yet to demonstrate that pre-training aids PassSum in the process of convergence on the code summarizing task. \figurename~\ref{figure:loss_curves} depicts the learning curves (containing the loss curve on the training set and the BLEU curve on the validation set) of the pre-trained PassSum and the randomly initialized PassSum on the code summarization task.

It can be seen from \figurename~\ref{figure:java_curves} and \figurename~\ref{figure:python_curves} that the pre-trained PassSum converges faster on the dataset with two code summaries compared to the randomly initialized PassSum. The loss, in particular, begins lower on the training set and eventually converges to a lower value, whereas the BLEU begins higher on the validation set and narrows to a higher score. Furthermore, both models perform early stops on both datasets, but the pre-training PassSum occurs earlier.

However, on the ruby dataset, we notice that the PassSum without pre-training has a smaller loss on the training set than the pre-trained PassSum, but the former has a lower BLEU on the validation set than the latter. We hypothesize that the PassSum without pre-training overfits on the training set due to the smaller ruby dataset, resulting in a drop in generalization capacity and consequently a worse result on the validation set.

So far, we can infer that pre-training not only allows PassSum to achieve greater performance and robustness, but it also allows PassSum to converge faster and achieve better performance in less time than PassSum without pre-training.

\subsubsection{RQ5: How effective is PassSum compared with the best baseline according to human evaluation?}
\label{section:rq5}

It is well known that automatic metrics (i.e., BLEU, METEOR and ROUGE-L) may not always accurately reflect the actual quality of the generated comments. Thus, we complement our automatic evaluation results with a human evaluation, where we manually evaluate the summaries generated by GraphCodeBERT (the best baseline in most cases) and PassSum. 

\textcolor{red}{To manually evaluate these summaries, we invite five participants who are not authors of this paper, they all have bachelor's degrees and are graduate students in software engineering. Besides, to ensure their ability of reading, understanding and documenting Java code, they are also required to meet the following criteria: (1) have more than four years of experience in Java programming, (2) have participated in the development of at least two Java software/projects, and (3) can understand and write English comment/document skillfully.}

Here is the human evaluation procedure. First, we randomly select 200 samples from the test set (including the ground-truth summaries). Then, we collect the summaries of these samples that were generated by GraphCodeBERT and PassSum. Finally, we ask the participants three questions:

\textit{Survey A: How useful is the summary?} In this survey, We first pair each of the 400 automatically generated comments with its ground truth (i.e., the human-written comment). Then we randomly divide these 400 pairs into 10 groups and ensure that each group contains 20 comments from GraphCodeBERT and 20 comments from PassSum, and that the comments generated by two models for a sample will not appear in the same group. The participants are told which comment in a pair is the ground truth and which one is automatically generated. They were asked to score each automatically generated comment against the gold standard on a scale of 0 to 5, \textcolor{red}{the scoring guide is shown below.}

\textbf{Score 0: crappy}, candidate and reference has no overlap in non-stop words and their meanings are completely different (no correct information).

\textbf{Score 1: useless}, candidate and reference has some overlap in non-stop words,but their meanings are completely different (no correct information)

\textbf{Score 2: mostly useless}, most of the summary contains wrong information but it does have some correct information.

\textbf{Score 3: somewhat useful}, candidate contains both useful and wrong information.

\textbf{Score 4: useful}, candidate contains useful information and has little or no wrong information, but some important information is missing.

\textbf{Score 5: very useful}, candidate contains all the important information in the reference and has no wrong information.

\begin{figure}[t]
    \centering
    \includegraphics[width=0.48\textwidth]{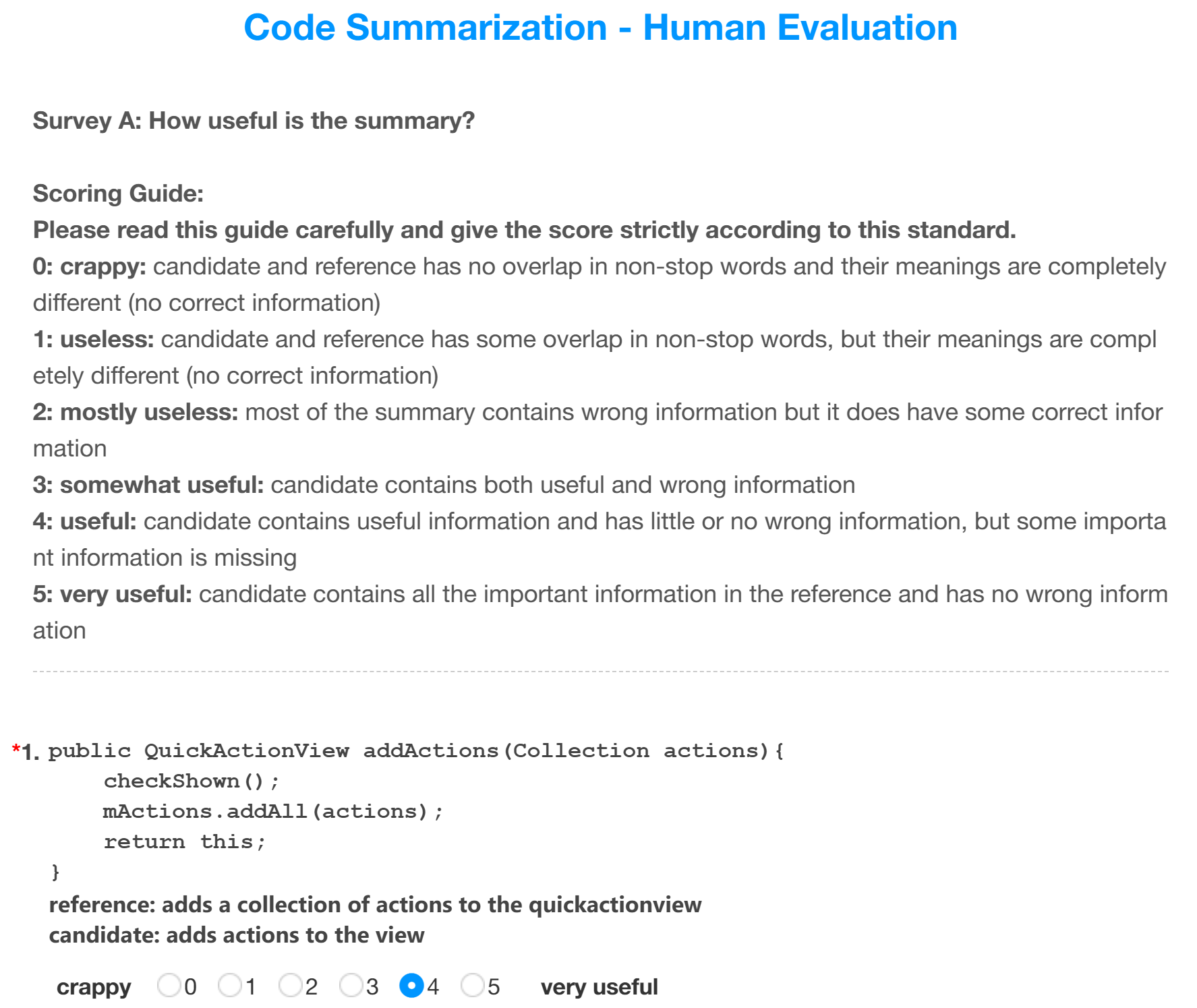}
    \caption{Details of the survey A.}
    \label{figure:survey_a}
\end{figure}

\figurename~\ref{figure:survey_a} shows the detailed description of survey A.

\begin{table}[t]
    \caption{Score distributions of two models in survey A.}
    \label{table:survey_a}
    \centering
    \resizebox{\linewidth}{!}{
    \begin{tabular}{lllllll}
        \toprule
        \textbf{}     & \textbf{0} & \textbf{1} & \textbf{2} & \textbf{3} & \textbf{4} & \textbf{5} \\
        GraphCodeBERT & 11.5\%     & 12\%       & 9\%        & 22.25\%    & 27\%       & 18.25\%    \\ 
        PassSum       & 7\%        & 7.25\%     & 8.75\%     & 22.25\%    & 29.25\%    & 25.5\%     \\
        \bottomrule
    \end{tabular}
    }
\end{table}

\textit{Survey B: Which of the two candidates are more useful?} In this survey, we pair the ground truth for each of the 200 samples with the comments generated by the two models for the same sample, so that we have 200 sets, each of which is composed of three comments. These 200 samples are then randomly divided into 10 groups. The participants are told which comment in each set is the ground truth and which ones are automatically generated, but they are not told which of the automatically generated comments is produced by which model. Moreover, the participants cannot infer which comment is generated by which model based on the order in which they are presented to them (since the order is random). The participants are then asked to compare which of the two automatically generated comments is more useful in light of the ground truth, assigning a score of 0 if the first candidate is considerably worse than the second and a score of 4 if the second candidate is considerably worse than the first. \textcolor{red}{The scoring criteria is shown below.}

\textbf{Score 0: much worse}, the first summary is a lot worse than the second summary.

\textbf{Score 1: somewhat worse}, the first summary is a bit worse than the second summary.

\textbf{Score 2: same}, none is better than the other.

\textbf{Score 3: somewhat better}, the first summary is a bit better than the second summary.

\textbf{Score 4: much better}, the first summary is a lot better than the second summary.

\begin{figure}[t]
    \centering
    \includegraphics[width=0.48\textwidth]{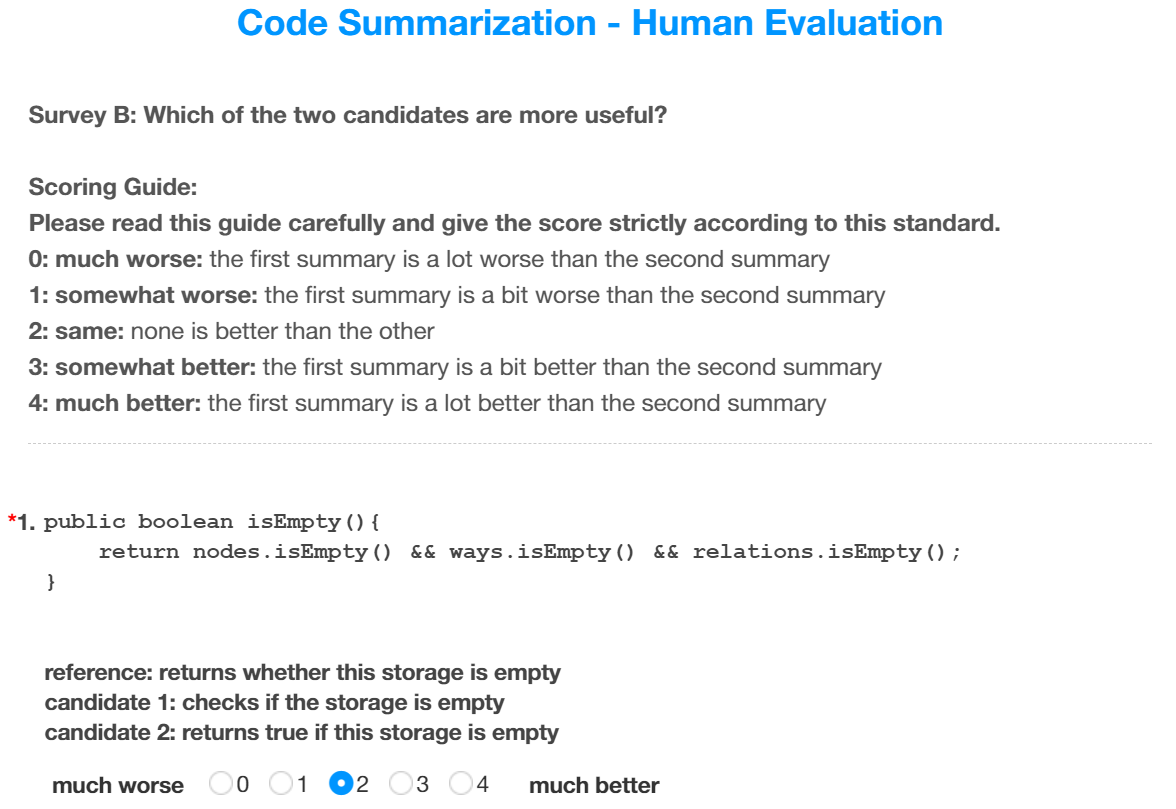}
    \caption{Details of the survey B.}
    \label{figure:survey_b}
\end{figure}

The illustration of survey B is shown in \figurename~\ref{figure:survey_b}.

Each participant is asked to evaluate 4 groups in each survey, so each group is evaluated by 2 different participants. Table~\ref{table:survey_a} shows the score distributions of the two models w.r.t. survey A. As we can see, PassSum gets more high scores (i.e., score$\geqslant$4) and less low scores (i.e., score$\leqslant$2) than GraphCodeBERT. Moreover, the average score of PassSum (3.36) is higher than that of GraphCodeBERT (2.96). These results provide suggestive evidence that PassSum performs better than GraphCodeBERT w.r.t. this survey. Table~\ref{table:survey_b} expresses the results of survey B in percentages. Specifically, after collecting the results from the participants, we tabulate the number of times PassSum produces a comment that is far better, slightly better, more or less the same, slightly worse, and far worse than GraphCodeBERT, using 4 to mean that PassSum is far better and 0 to mean that GraphCodeBERT is far better. As we can see from Table~\ref{table:survey_b}, there are more high scores in the result, which suggests that the comments generated by our model are perceived by the participants to be more useful than those generated by GraphCodeBERT. 

\begin{table}[t]
    \caption{Score distributions in survey B.}
    \label{table:survey_b}
    \centering
    \small
    \begin{tabular}{llllll}
        \toprule
                       & \textbf{0} & \textbf{1} & \textbf{2} & \textbf{3} & \textbf{4} \\
        percentage     & 3.25\%     & 10.5\%     & 39.25\%    & 28\%       & 18\%       \\
        \bottomrule
    \end{tabular}
\end{table}

\textit{Survey C: How good is the summary generated by PassSum compared to the ground truth?} In this survey, we pair the ground truth for the 200 samples with the comments generated by PassSum. These 200 pairs are then randomly divided into 10 groups. The participants are not told which one in the pair is generated by the model and which one is the ground truth. Moreover, since the two sentences in each pair is randomly shuffled, the participants cannot infer such information. The participants are then asked to compare the two sentences to see if they are similar, or which one is better, assigning a score of 0 if the first sentence is considerably better than the second, a score of 2 if these two sentence are almost identical, and a score of 4 if the second candidate is considerably better than the first. \textcolor{red}{Following is the scoring guideline.}

\textbf{Score 0: sentence 1 is absolutely better}, sentence 1 accurately and completely describes the function of the code, while sentence 2 does not reflect the function of the code at all.

\textbf{Score 1: sentence 1 is somehow better}, sentence 1 describes the function of the code more accurately than sentence 2, however, sentence 2 also partially describes the function of the code.

\textbf{Score 2: mostly identical}, the two sentences achieve almost the same effect in describing the function of the code.

\textbf{Score 3: sentence 2 is somehow better}, sentence 2 describes the function of the code more accurately than sentence 1, however, sentence 1 also partially describes the function of the code.

\textbf{Score 4: sentence 2 is absolutely better}, sentence 2 accurately and completely describes the function of the code, while sentence 1 does not reflect the function of the code at all.

\begin{figure}[t]
    \centering
    \includegraphics[width=0.48\textwidth]{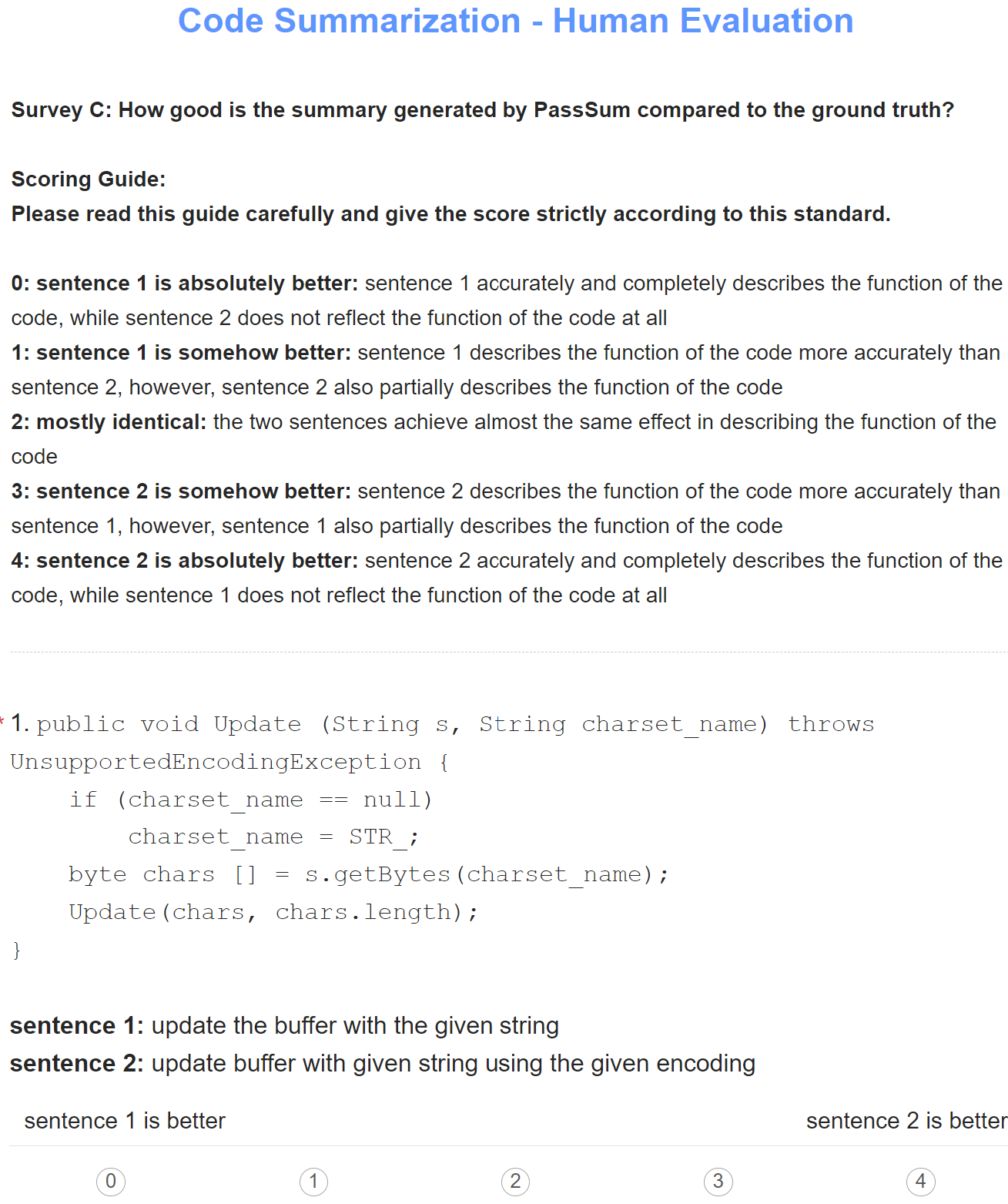}
    \caption{Details of the survey C.}
    \label{figure:survey_c}
\end{figure}

The illustration of survey C is shown in \figurename~\ref{figure:survey_c}.

Each participant is asked to evaluate 4 groups in each survey, so each group is evaluated by 2 different participants. After the survey, we recover the order, and the distribution of statistical scores is shown in Table~\ref{table:survey_c}, where we take 0 as the case where the ground truth is better than the sentences generated by PassSum, and 5 as the case where the sentences generated by PassSum are better. Table~\ref{table:survey_c} shows the score distributions of survey C.

\begin{table}[t]
    \caption{Score distributions in survey C.}
    \label{table:survey_c}
    \centering
    \small
    \begin{tabular}{llllll}
        \toprule
                       & \textbf{0} & \textbf{1} & \textbf{2} & \textbf{3} & \textbf{4} \\
        percentage     & 6.25\%     & 28.75\%     & 49.25\%    & 14\%       & 1.75\%       \\
        \bottomrule
    \end{tabular}
\end{table}

In almost half of the cases, the summaries generated by PassSum and the ground truth are essentially the same. In 6.25\% of cases, PassSum generates summaries that do not contain any code functionality at all. Conversely, PassSum also has examples that are more reflective of code functionality than ground truth, and these examples account for 15.75\% of the total. In summary, we can conclude that PassSum generates almost the same summary as ground truth with a general probability, however there are still a small number of cases that do not give a useful summary.

\subsection{Qualitative Analysis}
\label{section:qualitative_analysis}

With the human evaluation in Section~\ref{section:rq5}, we randomly select the results generated by PassSum and other models for 200 examples on the test set. In order to determine where PassSum outperforms other models and where PassSum fails, we revisit the summaries given by each model on these samples.

Through careful observation and summarization, we find that PassSum outperforms other baselines when there are more control components (e.g., \textit{if}, \textit{for} statements) in the code, especially when they are nested within each other and the code length is long. Conversely, when the code contains few or no control components, PassSum's performance is usually comparable to that of other models. We believe that the reason for this observation is that for examples containing more control information, their structural information is more important to understand the code, because it is difficult for the model to extract useful structural information directly from the code sequence only. And PassSum can extract useful structural information more accurately, which we believe is due to two factors: (1) Path embedding, which enables PassSum to obtain structural information from the input more easily outside the code sequence; (2) pre-training tasks, especially POP, train PassSum to process structural information in advance, making PassSum more aware of the structural information expressed through Path embedding.

% Through careful observation and summarization, we find that PassSum usually performs better than other models for examples with longer length, while for shorter length examples, PassSum usually performs comparable to or worse than other models. We believe that the reason for this observation is that for examples with longer lengths, the ability of the model to extract features from code sequences becomes lower, and therefore additional structural information is needed to bring more information. In contrast, for methods that encode structural information, Path Embedding has a shorter input length, so it can maintain a shorter input length for a large AST size, making the information density in the input larger and therefore allowing PassSum to extract more useful information. %这种长度的说法太普通了，感觉没有什么吸引力，再尝试挖掘比较新颖的说法。例如，“对于需要从代码逻辑中推理出新词的情况下，PassSum做的更好”？

We also give a few cases to demonstrate where PassSum outperforms baselines and where PassSum fails. Given a example, we use PassSum and three baselines (two pre-training models, CodeBERT and GraphCodeBERT, and one without pre-training, SiT) to predict the summary of the code. Results are given in Table~\ref{table:case_study}.

\begin{table*}
    \caption{Quantitative Analysis of PassSum and three baselines.}
    \label{table:quantitative_analysis}
    \centering
    \scriptsize
    \begin{tabular}{l|l}
        \toprule
            \multicolumn{1}{c|}{\textbf{Java}} &
            \multicolumn{1}{c}{\textbf{Python}} \\
        \midrule
            \begin{minipage}{0.5\textwidth}
            \begin{lstlisting}[language=Java]
public static float[] concatAllFloat(float[] ... arrays){
    int totalLength = _NUM;
    final int subArrayCount = arrays.length;
    for (int i = _NUM; i < subArrayCount; ++i)
        totalLength += arrays[i].length;
    float[] result = Arrays.copyOf(arrays[_NUM], totalLength);
    int offset = arrays[_NUM].length;
    for (int i = _NUM; i < subArrayCount; ++i){
        System.arraycopy(arrays[i], _NUM, result, offset, arrays[i].length);
        offset += arrays[i].length;
    }
    return result;
}
            \end{lstlisting}
            \end{minipage} &
            \begin{minipage}{0.5\textwidth}
            \begin{lstlisting}[language=Python,linewidth=0.95\textwidth]
def check_path_traversal(path, user='root', skip_perm_errors=False):
    for tpath in list_path_traversal(path):
        if (not os.access(tpath, os.R_OK)):
            msg = 'Could not access {0}.'.format(tpath)
            if (not os.path.exists(tpath)):
                msg += 'Path does not exist.'
            else:
                current_user = salt.utils.get_user()
                if (user != current_user):
                    msg += 'Try running as user {0}.'.format(user)
                else:
                    msg += 'Please give {0} read permissions.'.format(user)
            if skip_perm_errors:
                return
            raise SaltClientError(msg)

            \end{lstlisting}
            \end{minipage} \\
        \hline
            {\color{teal} CodeBERT}: concatenate all float arrays into an array &
            {\color{teal} CodeBERT}: traversal path and check access \\
            {\color{cyan} GraphCodeBERT}: concatenates all floats into an array &
            {\color{cyan} GraphCodeBERT}: traversal the path and check the access of user \\
            {\color{orange} SiT}: concats all float arrays into an array &
            {\color{orange} SiT}: list the given path and check if has access \\
        \hline
            {\color{purple} PassSum}: concatenates a list of float arrays into an array &
            \begin{tabular}[l]{@{}l@{}}
                {\color{purple} PassSum}: traverses the given path and checks whether the user has access to each\\path
            \end{tabular} \\
        \hline
            {\color{red} Human Written}: concatenates a list of float arrays into a single array &
            \begin{tabular}[l]{@{}l@{}}
                {\color{red} Human Written}: walk from the root up to a directory and verify that the current user\\has access to read each directory
            \end{tabular} \\
        \midrule
            \begin{minipage}{0.5\textwidth}
            \begin{lstlisting}[language=Java]
private void evaluatePopup(MouseEvent e){
    if (e.isPopupTrigger()){
        JPopupMenu menu = createBookmarkPopupMenu();
        if (menu != null){
            menu.show(this, e.getX(), e.getY());
        }
    }
}
            \end{lstlisting}
            \end{minipage} &
            \begin{minipage}{0.5\textwidth}
            \begin{lstlisting}[language=Python]
def temp_file_for(path):
    ext = os.path.splitext(path)[1]
    with NamedTemporaryFile(suffix=util.py3_path(ext), delete=False) as f:
        return util.bytestring_path(f.name)
            \end{lstlisting}
            \end{minipage} \\
        \hline
            {\color{teal} CodeBERT}: shows a popup menu &
            {\color{teal} CodeBERT}: create a temporary file as path name \\
            {\color{cyan} GraphCodeBERT}: creates a popup menu if necessary &
            {\color{cyan} GraphCodeBERT}: create a temp file with path \\
            {\color{orange} SiT}: checks the mouse event and popup a menu &
            {\color{orange} SiT}: create a temporary path \\
        \hline
            {\color{purple} PassSum}: shows a popup menu depending on the mouse event &
            {\color{purple} PassSum}: return a temporary file for given path \\
        \hline
            \begin{tabular}[l]{@{}l@{}}
                {\color{red} Human Written}: checks if the given mouse event is a popup trigger and\\creates a new popup menu if necessary
            \end{tabular} &
            \begin{tabular}[l]{@{}l@{}}
                {\color{red} Human Written}: return an unused filename with the same extension as the specified\\path
            \end{tabular} \\
        \bottomrule
    \end{tabular}
    \label{table:case_study}
\end{table*}

We list two examples in the Table~\ref{table:case_study} for each of the Java and Python languages. We consider the example at the top of the table to be a successful PassSum example, while the example at the bottom is a failed one. For the successful examples, it can be seen that although the summaries generated by PassSum are not exactly the same as those human written, they also accurately reproduce every functional detail. In the first Java example which contains many control components, the key information is embedded in the method name ``\texttt{concatAllFloat}'' and the result array (which corresponds to ``into a single array'' in the human written summary). PassSum accurately captures this key information and generates the correct summary. Meanwhile, other baselines also generate the correct summaries. But the situation becomes different in the first Python example, where the key information becomes more numerous, and most of the critical information is embedded in the control component. First there is ``\texttt{check}'' and ``\texttt{traversal}'' in the method name, then there is ``\texttt{user}'' in the parameters and body, and ``\texttt{access}'' in the fourth line. This time, only PassSum and GraphCodeBERT accurately captured this key information. However, for ``check access'' function of this example, PassSum accurately outputs ``checks whether the user has access to \textbf{\textit{each}} path'' with the captured control information (i.e., ``for'' statement), while GraphCodeBERT does not. In addition, the sentences generated by PassSum were more fluent and understandable, while the sentences generated by GraphCodeBERT has some errors and ambiguities, such as ``traversal'' being a noun, and making it difficult to understand whether the method checks the access of the given path, or the access of each path traversed. We can conclude from the above two examples that PassSum captures the key information in the code accurately and completely. We believe that it is Path Embedding that allows the model to get all the key information, especially for the control information, and thus generate accurate summaries. In addition, pre-training takes PassSum's understanding of the structural information of code even further, allowing it to accurately capture useful information from the input.

Below are two examples of PassSum failures. Just like baselines, PassSum ignores some details in both examples. One of the reasons for these failures is the methods are too short for PassSum to extract detailed descriptions, e.g., ``isPopupTrigger'' in the Java example and ``f.name'' in the Python example. However, owing to the ability of interpreting control component, PassSum notices the existence of the ``if'' condition and successfully generates ``depending on'' for the Java example, which does better than the other approaches. Another reason for all approaches failing in handling the Python example is that the given Python code does not expose key meaningful semantics through variable names and method names, i.e., more information are needed for the model to understand what the ``NamedTemporaryFile'' API does.

Through quantitative analysis, we can find that PassSum can accurately and completely capture the key information in the code through Path Embedding and pre-training, especially when the code is long and there is a lot of key information. However, there are occasions when some information is missed.

\section{Threats to Validity}
\label{section:threats}

\textit{Construct Validity.} Although we kept the other settings of all models as consistent as possible in our experiments, such as optimizer, initial learning rate, batch size, etc., we could not guarantee that the total number of parameters would be the same for all models due to the different neural networks of each model, especially in Table~\ref{table:exp_results}, instead we set the hidden layer size to the same size for all models, as we mentioned in Section~\ref{section:settings}.

In addition, due to the different input representation and processing approach, each model in Tables~\ref{table:exp_input} and Table~\ref{table:exp_time_space} has a slight difference in the total number of parameters. For example, the four models of multi-encoder have the most parameters due to two encoders they have, and the remaining models have the following size relationship in the total number of parameters: PassSum SBT (600)~\textgreater~PassSum code2seq (200)~\textgreater~PassSum~\textgreater~PassSum code2seq~\textgreater~PassSum SBT~\textgreater~PassSum None.

\textit{Internal Validity.} It is widely agreed that hyperparameters have a significant impact on the performance of deep learning models, but hyperparameters of PassSum are not tuned experimentally and are set empirically. Therefore, other hyperparameter settings may yield better results.

\textit{External Validity.} Many public datasets are proposed to explore code summarization. We select two widely used ones to perform experiments, but they may not be representative of other programming languages.

\section{Related Work}
\label{section:related_work}

There is a lot of proposed work relevant to this paper. First, there has been a lot of work done on the software engineering task that PassSum addresses, namely code summarization. Second, for the second and third pre-training objectives proposed in this paper, namely POP and MNG (see Section~\ref{section:pre_train_pop} and Section~\ref{section:pre_train_mng}), unlike the direct use of pre-training objectives present in NLP, these two are pre-training objectives that we designed specifically for source code. Similarly, there is also work proposing pre-training objectives designed for codes (not limited to application to the code summarization task). These tasks are new pre-training tasks designed for the properties that distinguish code from natural language. In this section, we mainly review some representative related efforts in the above two aspects.

\subsection{Code Summarization}
\label{section:related_cs}

Although code summarization was initially addressed through IR methods~\cite{marcus2003recovering,panichella2012mining,haiduc2010use,movshovitz2013natural,wong2013autocomment,allamanis2015bimodal,oda2015learning}, in recent years, neural network-based approaches have achieved a leading position.

CODE-NN~\cite{iyer2016summarizing} is the first data-driven code summarization model which employs LSTM networks with attention to produce sentences that describe C\# code snippets and SQL queries. DeepCom~\cite{hu2018deep}, an attention-based Seq2Seq model, convert ASTs to specially formatted sequences using a new structure-based traversal (SBT) method. Then input the AST sequences into the model to generate comments for Java methods. Wan et al.~\cite{wan2018improving} encodes the structure and sequential content of code via AST-based LSTM and LSTM respectively, then feeds the integrated code representation vector into a deep reinforcement learning framework, named actor-critic network. TL-CodeSum~\cite{hu2018summarizing} generates summaries by capturing semantics from the source code with the assistance of API knowledge, which is transferred into TL-CodeSum from API sequence summarization task. Code2seq~\cite{alon2018code2seq} is an alternative approach that represents a code snippet as the set of compositional paths in its AST and uses attention to select the relevant paths while decoding. Wei et al.~\cite{wei2019code} treat code summary as a dual task of code generation and propose a dual training framework to train the two tasks simultaneously. ASTNN~\cite{zhang2019novel} splits each large AST into a sequence of small statement trees, and encodes the statement trees to vectors by using a RvNN. Then a bidirectional RNN and a pooling layer is used to produce the vector representation of a code fragment. Rencos~\cite{zhang2020retrieval} combines IR and neural network methods. Firstly, it trains an attentional seq2seq model. During testing, for each sample, Rencos generate summaries by the fusion of retrieved code snippets and itself. It employs two code retrieval methods from the aspects of syntax and semantics to accommodate more knowledge about the code. Both ast-attendgru~\cite{leclair2019neural} and Hybrid-DeepCom~\cite{hu2020deep} combine words from code with code structure from an AST, the model processes each data source as a separate input, which allows the model to learn code structure independent of the text in code. Code+gnn+GRU~\cite{leclair2020improved} utilizes a sequence of source code tokens along with ConvGNNs to encode the AST of a Java method and generate natural language summaries.

With the introduction of Transformer~\cite{vaswani2017attention}, researchers use Transformer to further improve the performance of code summarization. Ahmad et al.~\cite{ahmad2020transformer} employ a Transformer-based model by incorporating relative position representations and copy attention~\cite{see2017get}. SiT~\cite{wu2021code} presents a novel structured-induced Transformer model on code summarization task. The model effectively incorporate multi-view structure into attention mechanism, and aggregate both global self-attention and structure-induced self-attention representations. MMTrans~\cite{yang2021multi} leverages the two modalities of the AST (i.e., SBT sequences and graphs) to represent both global and local semantic information of source code, then employs the two encoders and a joint decoder with the multi-head attention structure to capture the long-range dependencies between code tokens. BASTS~\cite{lin2021improving} uses block-wise code splitting and localizes the hierarchical syntax structure of AST generated for each code split. BASTS further adopts a pre-training strategy to capture local non-linear syntax encoding of split ASTs, which helps its Transformer-based summarization model generate high-quality code summaries.

\subsection{Pre-training Tasks for Source Code}
\label{section:related_tasks}

There is work on new pre-training tasks designed specifically for code, other than directly using pre-training tasks designed directly for natural language in NLP. To encourage the model to learn structure-aware representation, Guo et al.~\cite{guo2020graphcodebert} introduce Edge Prediction, which randomly samples 20\% of nodes in data flow and masks direct edges connecting these sampled nodes, and then predict these masked edges. In addition, to align representation between source code and data flow, they also introduce Node Alignment, which makes the model to predict edges between code tokens and nodes in data flow. Lachaux et al.~\cite{lachaux2021dobf} introduce a new pre-training objective, which obfuscates code snippets by replacing class, function and variable names with special tokens, and trains a model to recover the original names. When an identifier is selected, all of its instances in the code are replaced by the same special token. Similar to NSP, Next Code Segment Predicting~\cite{liu2020multi} is to predict whether two segments of code tokens follow each other in a piece of code snippet. Jain et al.~\cite{jain2020contrastive} hypothesize that programs with the same functionality should have similar underlying representations for downstream code understanding tasks. Therefore, they generate syntactically diverse but functionally similar programs with source-to-source compiler transformation techniques. Then, they employ contrastive objective to maximizes the similarity of these functionally similar programs.

\section{Conclusion}
\label{section:conclusion}

In light of two key weaknesses in existing neural code summarizers (namely, inefficient encoding of ASTs and lack of robust performance in the face of a small training set), we proposed PassSum. To handle the inefficient AST encoding problem, our novel Path Embedding offered better performance than code2seq-like encoding and SBT. As the training set size decreases, pre-training allowed PassSum's performance to deteriorate less quickly than its \textit{no pre-training} counterparts. Finally, in order to employ Transformer as PassSum's underlying architecture, we modified it so that it can assume two sequences as input. Importantly, while in previous work Transformer failed to benefit from ASTs, our method for incorporating ASTs allowed them to improve PassSum's performance.

\section*{Acknowledgment}
This work was supported by the National Natural Science Foundation of China (61802167), Natural Science Foundation of Jiangsu Province, China (BK20201250), Cooperation Fund of Huawei-NJU Creative Laboratory for the Next Programming, and NSF award 2034508. 

\ifCLASSOPTIONcaptionsoff
  \newpage
\fi



\bibliographystyle{IEEEtran}
\bibliography{refs}


% \end{thebibliography}


\begin{IEEEbiography}[{\includegraphics[width=1in,height=1.25in,clip,keepaspectratio]{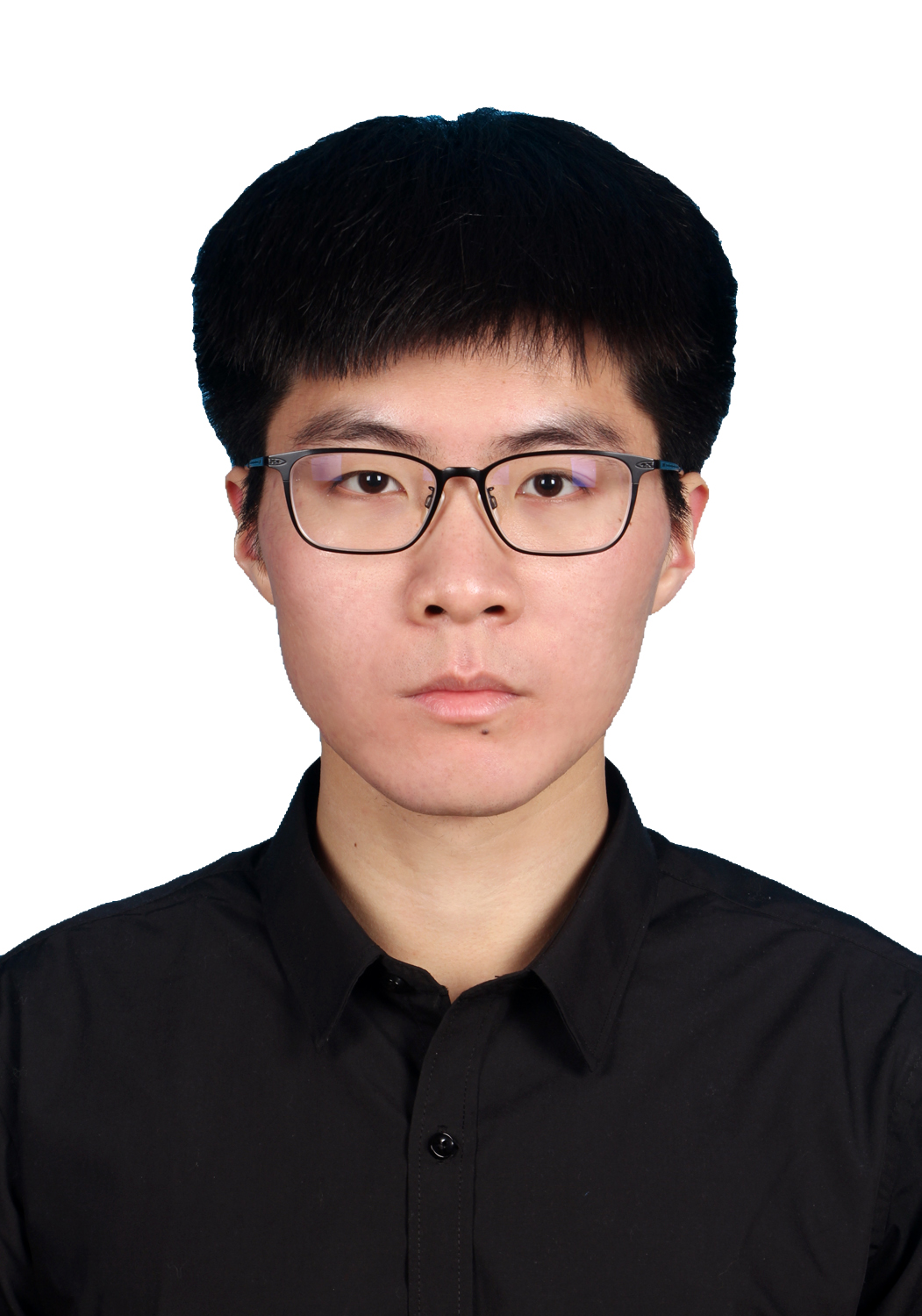}}]{Changan Niu}
received the MA.Eng. degree from Software Engineering, Nanjing University in 2019. He is currently working towards the PhD degree in Software Institute, Nanjing University. His research interests include software engineering, programming language processing and deep learning.
\end{IEEEbiography}

\begin{IEEEbiography}[{\includegraphics[width=1in,height=1.25in,clip,keepaspectratio]{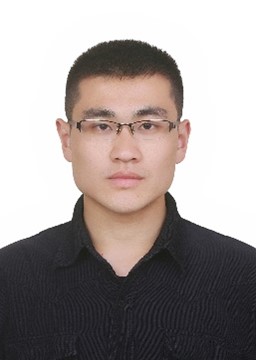}}]{Chuanyi Li}
received the B.S. degree and Ph.D. degree from Software Institute, Nanjing University in 2012 and 2017. He is a research assistant in Software Institute, Nanjing University. His research interests include software engineering, natural language processing, and process mining.
\end{IEEEbiography}
\vfill
\begin{IEEEbiography}[{\includegraphics[width=1in,height=1.25in,clip,keepaspectratio]{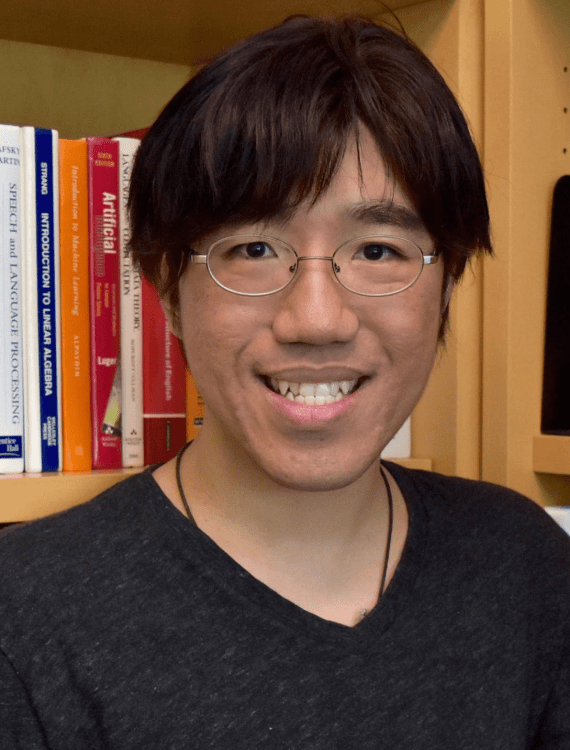}}]{Vincent Ng}
received the Ph.D. degree from CS department of Cornell University in 2004. He is a professor and doctoral supervisor in Computer Science, University of Texas at Dallas. His research interests include machine learning of natural language and software engineering.
\end{IEEEbiography}

\begin{IEEEbiography}[{\includegraphics[width=1in,height=1.25in,clip,keepaspectratio]{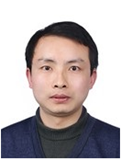}}]{Jidong Ge}
received the Ph.D. degree from CS department of Nanjing University in 2007. He is an associate professor in Software Institute, Nanjing University. His research interests include natural language processing, text mining, business process management and process mining.
\end{IEEEbiography}

\begin{IEEEbiography}[{\includegraphics[width=1in,height=1.25in,clip,keepaspectratio]{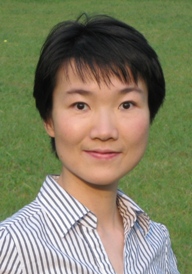}}]{Liguo Huang}
received her M.S degree and Ph.D. degree from the Computer Science Department at the University of Southern California in 2001 and December 2006. She is an associate professor in the Department of Computer Science at the Southern Methodist University. Her research interests include software quality/information dependability and value-based software engineering.
\end{IEEEbiography}

\begin{IEEEbiography}[{\includegraphics[width=1in,height=1.25in,clip,keepaspectratio]{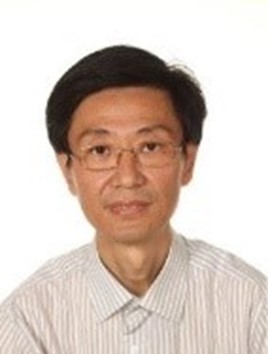}}]{Bin Luo}
is a professor, doctoral supervisor and associate dean of Software Institute, Nanjing University. His research interests include software engineering, workflow, process mining and text mining.
\end{IEEEbiography}
\vfill


%\newpage



%\vfill

% \enlargethispage{-5in}

\end{document}